\def \showauthors {Show authors}

\documentclass[letterpaper,twocolumn,10pt]{article}
\usepackage{usenix2019_v3}

\usepackage{array}
\usepackage{amsmath}
\usepackage{amssymb}
\usepackage{balance}
\usepackage{booktabs}
\usepackage{cite}
\usepackage{longtable}
\usepackage{multirow}
\usepackage{paralist}
\usepackage{subcaption}
\usepackage{tabularx}
\usepackage{url}
\usepackage{xcolor}
\usepackage{xspace}

\usepackage[underline=false]{pgf-umlsd}
\usetikzlibrary{positioning}

\newcommand{\sys}{\texttt{\sf{CACTI}}\xspace}

\newcommand{\tee}{\emph{TEE}\xspace}

\newcommand{\sysglobal}{\texttt{\sys-GLOBAL}\xspace}
\newcommand{\cl}{\emph{C}\xspace}
\newcommand{\sv}{\emph{S}\xspace}
\newcommand{\pa}{\emph{PA}\xspace}
\newcommand{\rp}{\emph{RP}\xspace}
\newcommand{\thr}{\emph{Th}\xspace}
\newcommand{\initphase}{Initialization Phase\xspace}

\newcommand{\ecall}{\texttt{ECALL}\xspace}

\ifdefined\showcomments
\usepackage{todonotes}
\usepackage{letltxmacro}
\LetLtxMacro{\todonote}{\todo}
\renewcommand{\todo}[2][]
{\todonote[inline, caption={#2}, size=\footnotesize, #1]
{\renewcommand{\baselinestretch}{0.5}\selectfont#2\par}}
\else
\usepackage[disable]{todonotes}
\fi
\ifdefined\showchanges
\newcommand{\changed}[1]{\color{blue}{#1}\color{black}\xspace}
\else
\newcommand{\changed}[1]{#1\xspace}
\fi
\ifdefined\showauthors
\author{
{\rm Yoshimichi Nakatsuka}\thanks{The first and second authors contributed equally to this work.}\\
{\normalsize University of California,}\\[-1mm] {\normalsize Irvine}
\\ {\small nakatsuy@uci.edu}
\and
{\rm Ercan Ozturk}\footnotemark[1]\\
{\normalsize University of California,}\\[-1mm] {\normalsize Irvine}
\\ {\small ercano@uci.edu}
\and
{\rm Andrew Paverd}\thanks{Work partially done while visiting the University of California, Irvine, as a US-UK Fulbright Cyber Security Scholar.}\\
{\normalsize Microsoft Research}
\\ {\small andrew.paverd@microsoft.com}
\and
{\rm Gene Tsudik}\\
{\normalsize University of California,}\\[-1mm] {\normalsize Irvine}
\\ {\small gene.tsudik@uci.edu}
}
\fi

\begin{document}
\ifdefined\showchanges
\onecolumn
\input{content/99-changes.tex}
\twocolumn
\fi

\title{\sys: Captcha Avoidance via Client-side TEE Integration}

\maketitle

\begin{abstract}

\changed{Preventing abuse of web services by bots} is an increasingly important problem, as abusive activities 
grow in both volume and variety. CAPTCHAs are the most common way for thwarting bot activities. 
However, they are often ineffective against bots and frustrating for humans.
In addition, some recent CAPTCHA techniques diminish user privacy.
Meanwhile, client-side Trusted Execution Environments (TEEs) are becoming increasingly widespread 
(notably, ARM TrustZone and Intel SGX), allowing establishment of trust in a small part (trust anchor or TCB) of client-side hardware.
This prompts the question: can a TEE help reduce (or remove  entirely) user burden of solving CAPTCHAs?

In this paper, we design \sys: {\bf\underline C}APTCHA {\bf\underline A}voidance via {\bf\underline C}lient-side {\bf\underline T}EE {\bf\underline I}ntegration. 
Using client-side TEEs, \sys allows legitimate clients to generate unforgeable \emph{rate-proofs} demonstrating 
how frequently they have performed specific actions.
These rate-proofs can be sent to web servers in lieu of solving CAPTCHAs.
\sys provides strong client privacy guarantees, since the information is only sent to the visited 
website and authenticated using a group signature scheme.
Our evaluations show that overall latency of generating and verifying a \sys rate-proof is less than $0.25$ sec, 
while \sys's bandwidth overhead is over $98\%$ lower than that of current CAPTCHA systems.

\end{abstract}

\section{Introduction}
\label{sec:intro}
\begin{figure}[t] 
    \centering
    \begin{subfigure}{0.47\textwidth}
    	\centering
    	\includegraphics[width=0.55\linewidth]{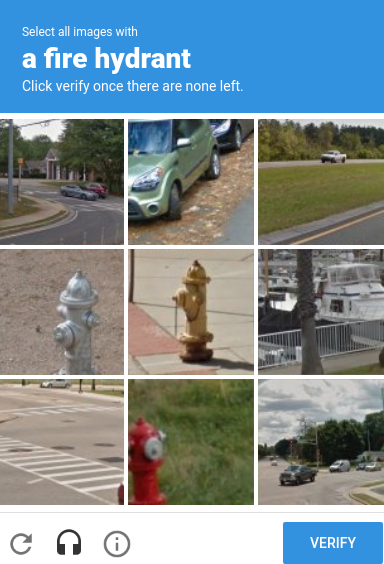}
    	\caption{Image-based object recognition reCAPTCHA~\cite{recaptcha}}
    	\label{fig:visualcaptcha}
    \end{subfigure}        
    \begin{subfigure}{0.47\textwidth}
        \centering
        \includegraphics[width=0.7\linewidth]{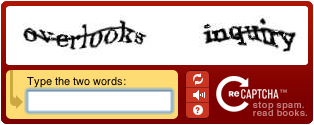}
        \caption{Image-based text recognition reCAPTCHA~\cite{recaptcha}}
        \label{fig:captcha-text}
    \end{subfigure}
    \begin{subfigure}{0.47\textwidth}
        \centering
        \includegraphics[width=0.6\linewidth]{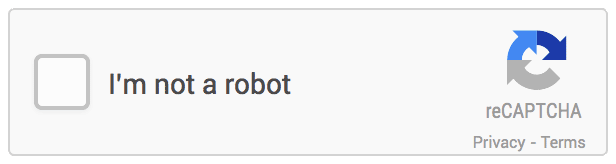}
        \caption{Behavior-based reCAPTCHA~\cite{recaptcha}}
        \label{fig:recaptcha}
    \end{subfigure}
    \caption{Examples of CAPTCHAs}
    \label{fig:captchas}
\end{figure}

In the past two decades, as Web use became almost universal and abuse of Web services grew dramatically, there has been an increasing trend (and real need) to use security tools that help \changed{prevent abuse by automated means, i.e., so-called {\bf bots}}.
The most popular mechanism is CAPTCHAs: Completely Automated Public Turing test to tell Computers and Humans Apart~\cite{VonAhn2003}.
A CAPTCHA is essentially a puzzle, such as \changed{an object classification task (Figure~\ref{fig:visualcaptcha}) or distorted text recognition (see Figure~\ref{fig:captcha-text})}, that aims to confound (or at least slow down) a bot, while being \changed{easily\footnote{Exactly what it means to be ``easily''  solvable is subject to some debate.}} solvable by a human user.
CAPTCHAs are often used to protect sensitive actions, such as creating a new account or submitting a web form.

\changed{
Although primarily intended to distinguish humans from bots, it has been shown that CAPTCHAs are not very effective at this task~\cite{motoyama2010re}.
\changed{Many CAPTCHAs can be solved by algorithms (e.g., image recognition software) or outsourced to human-driven \emph{CAPTCHA-farms}\footnote{A CAPTCHA farm is usually sweatshop-like operation, where employees solve CAPTCHAs for a living.}
to be solved on behalf of bots.}
Nevertheless, CAPTCHAs are still widely used to \changed{increase the adversary's costs (in terms of time and/or money) and} reduce the \emph{rate} at which bots can perform sensitive actions.
For example, computer vision algorithms are computationally expensive, and outsourcing to CAPTCHA-farms costs money and takes time.
}

From the users' perspective, CAPTCHAs are generally unloved (if not outright hated), since they represent a barrier and an annoyance (aka denial-of-service) for legitimate users. 
Another major issue is that most CAPTCHAs are visual in nature, requiring sufficient ambient light and screen resolution, as well as good eyesight.
Much less popular audio CAPTCHAs are notoriously poor, and require a quiet setting, decent-quality audio output facilities, as well as good hearing. 

More recently, the reCAPTCHA approach has become popular. 
It aims to reduce user burden by having users click a checkbox (Figure~\ref{fig:recaptcha}), while performing behavioral analysis of the user's browser interactions.
Acknowledging that even this creates friction for users, the latest version (``invisible reCAPTCHA'') does not require any user interaction.
However, the reCAPTCHA approach is potentially detrimental to {\bf user privacy} because it requires maintaining long-term state, e.g., in the form of Google-owned cookies.
\changed{Cloudflare recently decided to move away from reCAPTCHA due to privacy concerns and changes in Google's business model~\cite{cfhCaptcha}.}

Notably, all current CAPTCHA-like techniques are server-side, i.e., they do not rely on any security features of, or make any trust assumptions about, the client platform.
The purely server-side nature of CAPTCHAs was reasonable when client-side hardware security features were not widely available. 
However, this is rapidly changing with the increasing popularity of Trusted Execution Environments (TEEs) on a variety of computing platforms, e.g., TPM and Intel SGX for desktops/laptops and ARM TrustZone for smartphones and even smaller devices.
Thus, it is now realistic to consider abuse prevention methods that include client-side components.
For example, if a TEE has a trusted path to some form of user interface, such as a mouse, keyboard, or touchscreen, this \emph{trusted User Interface (UI)} could securely confirm user presence. 
Although this feature is still unavailable on most platforms, it is emerging through features like Android's Protected 
Confirmation~\cite{AndroidProtectedConfirmation}. 
This approach's main advantages are minimized user burden (e.g., just a mouse click) and increased security, since it would be impossible for software to forge this action. Admittedly however, this approach can be defeated by adversarial hardware e.g., a programmable USB peripheral that pretends to be a mouse or keyboard.

However, since the majority of consumer devices do not currently have a trusted UI, it would be highly desirable to reduce the need for CAPTCHAs using only existing TEE functionality.
\changed{As discussed above, the main goal of modern CAPTCHAs is to increase adversarial costs and reduce the \emph{rate} at which they can perform sensitive actions.}
Therefore, if legitimate users had a way to prove that their rate of performing sensitive actions is below some threshold, a website could decide to allow these users to proceed without solving a CAPTCHA.
If a user can not provide such a proof, the website could simply fall back to using CAPTCHAs.
Though this would not fully prevent bots, it would not give them any advantage compared to the current arrangement of using CAPTCHAs.

Motivated by the above discussion, this paper presents \sys{}, a flexible mechanism for allowing legitimate users to prove to websites that they are not acting in an abusive manner.
By leveraging widespread and increasing availability of client-side TEEs, \sys{} allows users to produce \emph{rate-proofs}, which can be presented to websites in lieu of solving CAPTCHAs.
A rate-proof is a simple assertion that:
\begin{compactenum} 
\item The rate at which a user has performed some action is below a certain threshold, and
\item The user's time-based counter for this action has been incremented.
\end{compactenum}
When serving a webpage, the server selects a \emph{threshold} value and sends it to the client.
If the client can produce a rate-proof for the given threshold, the server allows the action to proceed without showing a CAPTCHA. Otherwise, the server presents a CAPTCHA, as before. In essence, \sys{} can be seen as a type of ``express checkout'' for legitimate users.

One of the guiding principles and goals of \sys{} is user privacy -- it reveals only the minimum amount of information and sends this directly to the visited website. Another principle is that the mechanism should not mandate any specific security policy for websites. Websites can define their own security policies e.g., by specifying thresholds for rate-proofs.
Finally, \sys{} should be configurable to operate without any user interaction, in order to make it accessible to all users, including those with sight or hearing disabilities.
\changed{Although chiefly motivated by the shortcomings of CAPTCHAs, we believe that the general approach of client-side (TEE-based) rate-proofs, can also be used in other common web scenarios.}
For example, news websites could allow users to read a limited number of articles for free per month, without relying on client side cookies (which can be cleared) or forcing users to log-in (which is detrimental to privacy).
Online petition websites could check that users have not signed multiple times, without requiring users to provide their email addresses, which is once again, detrimental to privacy.
We therefore believe that our TEE-based rate-proof concept is a versatile and useful web security primitive.

Anticipated contributions of this work are:
\begin{compactenum}
\item We introduce the concept of a \emph{rate-proof}, a versatile web security primitive that allows legitimate users to securely prove that their rate of performing sensitive actions falls below a server-defined threshold.
\item We use the rate-proof as the basis for a concrete client-server protocol that allows legitimate users to present rate-proofs in lieu of solving CAPTCHAs.
\item We provide a proof-of-concept implementation of \sys, over Intel SGX, realized as a Google Chrome browser extension.
\item We present a comprehensive evaluation of security, latency, and deployability of \sys.
\end{compactenum}

\noindent{\bf Organization}:
Section~\ref{sec:bg} provides background information, and Section~\ref{sec:tmodreq} defines our threat model and security requirements.
Next, Section~\ref{sec:smoddchal} presents our overall design and highlights the main challenges in realizing this.
Then, Section~\ref{sec:impl} explains our proof of concept implementation and discusses how \sys overcomes the design challenges, followed by Section~\ref{sec:eval} which presents our evaluation of the security, performance, and deployability of \sys.
Section~\ref{sec:disc} discusses further optimizations and deployment considerations, and Section~\ref{sec:relw} summarizes related work.

\section{Background} \label{sec:bg}

\subsection{Trusted Execution Environments} \label{subsec:tee}
A Trusted Execution Environment (TEE) is a primitive that protects confidentiality and integrity of 
security-sensitive code and data from untrusted code.
A typical TEE provides the following features:

\textbf{Isolated execution.}
The principal function of a TEE is to provide an execution environment that is isolated from all other software 
on the platform, including privileged system software, such as the OS, hypervisor, or BIOS.
Specifically, data inside the TEE can only be accessed by the code running inside the TEE.
The code inside the TEE provides well-defined entry points (e.g., call gates), which are enforced by the TEE.

\textbf{Remote attestation.}
Remote attestation provides a remote party with strong assurances about the TEE and the code running therein.
Specifically, the TEE (i.e., the \emph{prover}) creates a cryptographic assertion that: (1) demonstrates 
that it is a genuine TEE, and (2) unambiguously describes the code running in the TEE.
The remote party (i.e., the \emph{verifier}) can use this to decide whether to trust the TEE, and 
then to bootstrap a secure communication channel with the TEE.

\textbf{Data sealing.}
Data sealing allows the code running inside the TEE to encrypt data such that it can be securely stored outside the TEE.
This is typically implemented by providing the TEE with a symmetric \emph{sealing key}, which can be used to 
encrypt/decrypt the data. In current TEEs, sealing keys are platform-specific, meaning that data can only be 
unsealed on the same platform on which it was sealed.

\textbf{Hardware monotonic counters.}
A well known attack against sealed data is \emph{rollback}, where the attacker replaces the sealed data 
with an older version.Mitigating this requires at least some amount of rollback-protected storage, typically 
realized as a hardware monotonic counter.
When sealing, the counter can be incremented and the latest value is included in the sealed data.
When unsealing, the TEE checks that the included value matches the current hardware counter value.
Since hardware counters themselves require rollback-protected storage, TEEs typically only have a small number of counters.

One prominent TEE example is \emph{Intel Software Guard Extensions (SGX)}~\cite{anati2013innovative, 
hoekstra2013using, mckeen2013innovative}.
SGX is a hardware-enforced TEE available on Intel CPUs from the Skylake microarchitecture onwards.
SGX allows applications to create isolated environments, called \emph{enclaves}, running in the application's virtual address space.
A special region in physical memory is reserved for enclaves, called the Enclave Page Cache (EPC). 
The EPC can hold up to 128MB of code and data, shared between all running enclaves.
When enclave data leaves the CPU boundary, it is transparently encrypted and integrity-protected by CPU's 
Memory Encryption Engine (MEE) to defend against physical bus snooping/tampering attacks.
Since enclaves run in the application's virtual address space, enclave code can access all the memory of its 
host application, even that outside the enclave.
Enclave code can only be called via predefined function calls, called \texttt{ECALL}s.

Every enclave has an enclave identity (\texttt{MRENCLAVE}), which is a cryptographic hash of the code that has been 
loaded into the enclave during initialization, and various other configuration details.
Each enclave binary must be signed by the developer, and the hash of the developer's public key is 
stored as the enclave's signer identity (\texttt{MRSIGNER}).

SGX provides two types of attestation: \changed{local and remote.
Local attestation allows two enclaves running on the same platform to confirm each other's identity and communicate 
securely, even though this communication goes via the untrusted OS. SGX uses local attestation to build  
remote attestation.} Specifically, an application enclave performs local attestation with an Intel-provided \emph{quoting enclave}, 
which holds a group private key provisioned by Intel.
The quoting enclave verifies the local attestation and creates a signed \emph{quote}, which includes the application 
\changed{enclave's and signer's identities}, as well as user-defined data provided by the application enclave.
This quote is sent to the remote verifier, which, in turn, uses the Intel Attestation Service (IAS) to verify it. 
Since the attestation uses a group signature scheme, the verifier cannot determine whether two quotes 
were generated by the same platform.

In SGX, data can be sealed in one of two modes, \changed{based on: (1) the enclave's identity, such that only the same 
type of enclave can unseal it, or (2) the signer identity, such that any enclave signed by the same developer 
(running on the same platform) can unseal it.}
SGX provides hardware monotonic counters and allows each enclave to use up to 256 counters \changed{at a time}.
\subsection{Group Signatures} \label{subsec:gs}
A group signature scheme aims to prevent the verifier from determining \changed{the group member which generated the signature.
Each group member} is assigned a group private key under a single group public key.
In case a group member needs to be revoked, a special entity called {\em group manager} can open the signature.
A group signature scheme \changed{is composed of  five algorithms~\cite{Ateniese2000group}:}
\begin{compactitem}
    \item \textbf{Setup:} Given a security parameter, an efficient algorithm outputs a group public key and a 
    master secret for the group manager.
    \item \textbf{Join:} A user interacts with the group manager to receive a 
    group private key and a membership certificate.
    \item \textbf{Sign:} Using the group public key, group private key, membership certificate, and a message \textit{m}, 
    a group member generates a group signature of \textit{m}.
    \item \textbf{Verify:} Using the group public key, an entity \changed{verifies a group signature}.
    \item \textbf{Open:} Given a message, \changed{a putative signature on the message, the group public key and the master secret, 
    the group manager determines the identity of the signer.}
\end{compactitem}
A secure group signature scheme satisfies the following properties~\cite{Ateniese2000group}:
\begin{compactitem}
    \item \textbf{Correctness:} Signatures \changed{generated with any member's group private key} must 
    be verifiable by the group public key.
    \item \textbf{Unforgeability:} \changed{Only an entity that holds a group private key can} generate signatures.
    \item \textbf{Anonymity:} Given a group signature, it must be computationally hard for anyone (except the group manager) 
    to identify the signer.
    \item \textbf{Unlinkability:} Given two signatures, it must be computationally hard to determine whether these 
    were signed by the same group member.
    \item \textbf{Exculpability:} Neither a group member nor the group manager can generate signatures on behalf of 
    other group members.
    \item \textbf{Traceability:} The group manager can determine the identity of a group member that generated a particular signature.
    \item \textbf{Coalition-resistance:} Group members cannot collude to create a signature that cannot be linked to one of the 
    group members by the group manager.
\end{compactitem}
Enhanced Privacy ID (EPID)~\cite{epid} is a group signature scheme used by remote attestation of Intel SGX enclaves.
It satisfies the above properties whilst providing additional privacy-preserving revocation mechanisms to 
revoke compromised \changed{or misbehaving} group members.
Specifically, EPID's \emph{signature-based revocation} protocol does not ``Open'' signatures
but rather uses a signature produced by the revoked member to notify other entities that this particular member has been revoked.

\section{System \& Threat Models} \label{sec:tmodreq}
The ecosystem that we consider includes three types of principals/players: (1) servers, (2) clients, and (3) TEEs.
There are multitudes of these three principal types.
The number of clients is the same as that of TEEs, and each client houses exactly one TEE.
Even though a TEE is assumed to be physically within a client, we consider it to be separate security entity.
Note that a human user can, of course, operate or own multiple clients, although there is clearly a limit 
and more clients implies higher costs for the user.

We assume that all TEEs are trusted: honest, benign and insubvertible.
We consider all side-channel and physical attacks against TEEs to be out of scope of this work and assume 
that all algorithms and cryptographic primitives implemented within TEEs are impervious to such attacks.
We also consider cuckoo attacks, whereby a malicious client utilizes multiple (possibly malware infected) machines 
with genuine TEEs, to be out of scope, since clients and their TEEs are not considered to be strongly \changed{bound.}
We refer to \cite{presenceattestation} and \cite{initpresenceattestation} {\changed as far as means for} countering such attacks.
We assume that servers have a means to authenticate and attest TEEs, possibly with the help of the TEE manufacturer.

All clients and servers are untrusted, i.e., they may act maliciously. The goal of a malicious client is to avoid 
CAPTCHAs, while a malicious server either aims to inconvenience a client (via DoS) or violate client's privacy.
For example, a malicious server can try to learn the client's identity or link multiple visits by the same client.
Also, multiple servers may collude in an attempt to track clients.

Our threat model yields the following requirements for the anticipated system:
\begin{compactitem}
	\item \textbf{Unforgeability:} Clients cannot forge or modify \sys{} rate-proofs.
	\item \textbf{Client privacy:} A server (or a group \changed{thereof}) cannot link rate-proofs to the clients 
	that generated them.
\end{compactitem}
We also pose the following non-security goals:
\begin{compactitem}
	\item \textbf{Latency:} User-perceived latency should be minimized.
	\item \textbf{Data transfer:} The amount of data transfer between client and server should be minimized.
	\item \textbf{Deployability:} The system should be deployable on current off-the-shelf client and server hardware.
\end{compactitem}

\section{\sys Design \& Challenges} \label{sec:smoddchal}
This section discusses the overall design of \sys and justifies \changed{our design choices}.

\subsection{Conceptual Design} \label{subsec:concept}
\textbf{Rate-proofs.}
The central concept underpinning our design is the \emph{rate-proof} (\rp). Conceptually, the idea is as follows:
Assuming that a client has an idealized TEE, the TEE stores one or more named sorted lists of timestamps in its rollback-protected secure memory.
To create a rate-proof for a specific list, the TEE is given the name of the list, a \emph{threshold} (\thr), and a new timestamp ($t$).
The threshold is expressed as a starting time ($t_s$) and a count ($k$).
This can be interpreted as: \emph{``no more than $k$ timestamps since $t_s$''}.
The TEE checks that the specified list contains $k$ or fewer timestamps with values greater than or equal to $t_s$.
If so, it checks if the new timestamp $t$ is greater than the latest timestamp in the list.
If both checks succeed, the TEE pre-pends $t$ to the list and produces a signed statement confirming that the named list is below the specified threshold and the new timestamp has been added.
If either check fails, no changes are made to the list and no proof is produced.
Note that the rate-proof does not disclose the number of timestamps in the list.

Furthermore, each list can also be associated with a public key.
In this case, requests for rate-proofs must be accompanied by a signature over the request that can be verified with the associated public key.
This allows the system to enforce a \emph{same-origin} policy for specific lists -- proofs over such lists can only be requested by the same entity that created them.
Note that this does not provide any binding to the \emph{identity} of the entity holding the private key, as doing so would necessitate the TEE to check identities against a global public key infrastructure (PKI) and we prefer for \sys not to require it.

\changed{
Rate-proofs differ from rate \emph{limits} because the user is allowed to perform the action any number of times. 
However, once the rate exceeds the specified threshold, the user will no longer be able to produce rate-proofs.
The client can always decide to \emph{not} use its TEE; this covers clients who do not have TEEs or those whose rates exceeded the threshold. On the other hand, if the server does not yet support \sys{}, the client does not store any timestamps, or perform any additional computation.
}

\textbf{CAPTCHA-avoidance.}
In today's CAPTCHA-protected services, the typical interaction between the client (\cl) and server (\sv) proceeds as follows: 
\begin{compactenum}
	\item \cl requests access to a service on \sv.
	\item \sv returns a CAPTCHA for \cl to solve.
	\item \cl submits the solution to \sv.
	\item If the solution is verified, \sv allows \cl access to the service.
\end{compactenum}
Although modern approaches, e.g., reCAPTCHA, might include additional steps (e.g., communicating with third-party services), these can be abstracted into the above pattern.

Our CAPTCHA-avoidance protocol keeps the same interaction sequence, while substituting steps~2 and~3 with rate-proofs.
Specifically, in step~2, the server sends a threshold rate and the current timestamp.
In step~3, instead of solving a CAPTCHA, the client generates a rate-proof with the specified threshold and timestamp, and submits it to the server.
The server has two types of lists:
\begin{compactitem}
\item \textbf{Server-specific:} The server requests a rate-proof over its own list.
The name of the list could be the server's URL, and the request may be signed by the server.
This determines the rate at which the client visits this specific server.
\item \textbf{Global:} The server requests a rate-proof over a global list, with a well-known name, e.g. \texttt{\sys{}-GLOBAL}.
This yields the rate at which the client visits all servers that use the global list.
\end{compactitem}
The main idea of CAPTCHA avoidance is that a legitimate client should be able to prove that its rate is below the server-defined threshold.
In other words, the server should have sufficient confidence that the client is not acting in an abusive manner (where the threshold of between abusive and non-abusive behaviors is set by the server).
Servers can select their own thresholds according to their own security requirements.
A given server can vary the threshold across different actions or even across different users or user groups, e.g., lower thresholds for suspected higher-risk users.
If a client cannot produce a rate-proof, or is unwilling to do so, the server simply reverts to the current approach of showing a CAPTCHA.
\changed{\sys essentially provides a \emph{fast-pass} for legitimate users.}

\changed{
The original CAPTCHA paper~\cite{VonAhn2003} suggested that CAPTCHAs could be used in the following scenarios:
\begin{compactenum} 
    \item \textbf{Online polls:} to prevent bots from voting,
    \item \textbf{Free email services:} to prevent bots from registering for thousands of accounts,
    \item \textbf{Search engine bots:} to preclude or inhibit indexing of websites by bots,
    \item \textbf{Worms and spam:} to ensure that emails are sent by humans,
    \item \textbf{Preventing dictionary attacks.} to limit the number of password attempts.
\end{compactenum}

As discussed in Section~\ref{sec:intro}, it is unrealistic to assume that CAPTCHAs cannot be solved by bots (e.g., using computer vision algorithms) or outsourced to CAPTCHA farms.
Therefore, we argue that all current uses of CAPTCHAs are actually intended to slow down attackers or increase their costs.
In the list above, scenarios 2 and 5 directly call for rate-limiting, while scenarios 1, 3, and 4 can be made less profitable for attackers if sufficiently rate-limited. Therefore, \sys can be used in all these scenarios.

In addition to CAPTCHAs, modern websites use a variety of abuse-prevention systems (e.g., filtering based on client IP address or cookies).
We envision \sys being used alongside such mechanisms.
Websites could dynamically adjust their \sys rate-proof thresholds based on information from these other mechanisms.
We are aware that rate-proofs are a versatile primitive that could be used to fight abusive activity in other ways, or even enable new use-cases.
However, in this paper, we focus on the important problem of reducing the user burden of CAPTCHAs.
}

\subsection{Design Challenges} \label{subsec:dchal}
In order to realize the conceptual design outlined above, we identify the following key challenges:

\textbf{TEE attestation.} In current TEEs, the process of remote attestation is not standardized. 
For example, in SGX, a verifier must first register with Intel Attestation Service (IAS) before it can verify TEE quotes.
Other types of TEEs would have different processes.
It is unrealistic to expect every web server to establish a relationship with the IAS and equivalent services from other manufacturers in order to verify attestation results.
Therefore, web servers cannot play the role of the attestation verifier.
However, they still need some way to determine that the client is running a genuine TEE.

\textbf{TEE memory limitations.} TEEs typically have a small amount of secure memory.
For example, if the memory of an SGX enclave exceeds the size of the EPC (usually 128~MB), the CPU has to swap pages out of the EPC.
This is a very expensive operation, since these pages must be encrypted and integrity protected.
Therefore, \sys should minimize the required amount of enclave memory, since other enclaves may be running on the same platform.
This is a challenge because we have to store timestamps indefinitely.

\textbf{Limited number of monotonic counters.} TEEs typically have a limited number of hardware monotonic counters, e.g., SGX allows at most 256 per enclave.
Also, the number of counter increments can be limited, e.g., in SGX the limit is 100 in a single epoch~\cite{mc} -- a platform power cycle, or a 24~hour period.
This is a challenge because hardware monotonic counters are critical for achieving rollback-protected storage.
Recall that \sys{} requires rollback-protected storage for all timestamps, to prevent malicious clients from rolling-back the timestamp lists and falsifying rate-proofs.
Furthermore, this storage must be updated every time a new timestamp is added, i.e., for each successful rate-proof.

\textbf{TEE entry/exit overhead.} Invoking TEE functionality typically incurs some overhead.
For example, whenever an execution thread enters/exits an SGX enclave, the CPU has to perform various checks and procedures (e.g., clearing registers) to ensure that enclave data does not leak.
Identifying and minimizing the number of TEE entries/exits, whilst maintaining functionality, can be challenging.

\subsection{Realizing \sys{} Design} \label{subsec:realizing}
We now present a detailed design that addresses aforementioned design challenges.
We describe its implementation in Section~\ref{sec:impl}.

\subsubsection{Communication protocol} \label{subsubsec:comm}
The web server must be able to determine that a supplied rate-proof was produced by a genuine TEE.
Typically, this would be done using remote attestation, where the TEE proves that it is running \sys{} code.
If the TEE provides privacy-preserving attestation (e.g., the EPID protocol used in SGX remote attestation), this would also fulfill our requirement for client privacy, since websites would not be able to link rate-proofs to specific TEEs.

However, as described above, current TEE remote attestation is not designed to be verified by anonymous third parties.
For example, with SGX, every web server operator would have to register with Intel to \changed{engage in remote attestation via IAS.} Furthermore, as \sys{} is not limited to any particular TEE type, websites would need to understand attestation results from multiple TEE vendors, potentially using different protocols.
Finally, some types of TEEs might not support privacy-preserving remote attestation, which would undermine our requirement for client privacy.

\changed{
To overcome this challenge, we introduce a separate \emph{Provisioning Authority} (\pa) in order to unify various processes for attesting \sys{} TEEs.
Fundamentally, the \pa is responsible for verifying TEE attestation (possibly, via the TEE vendor) and establishing a privacy-preserving mechanism through which websites can also establish trust in the TEE.
Specifically, the \pa protects user privacy by using the EPID group signature scheme.
The \pa plays the role of the EPID \emph{issuer}, and -- optionally -- the \emph{revocation manager}~\cite{epid}.
During the \initphase (as shown in Figure~\ref{fig:init_phase}), the \pa verifies the attestation from the client's TEE and then runs the EPID \texttt{join} protocol with the client's TEE in order to provision the TEE with a group private key $G_{sk_{\tee}}$.
The \pa certifies and publishes the group public key $G_{pk}$.
During this \initphase, the \pa may optionally require the client to prove their identity (e.g., by signing into an account) -- this is a business decision and different \pa{s} may take different approaches.
However, after the \initphase, the \pa cannot link signatures to any specific client thanks to the properties of the underling BBS+ signature scheme and signature-based revocation used in EPID~\cite{epid}.
We analyze security implications of malicious \pa{s} in Section~\ref{subsec:seceval}, and discuss the use of other group signature schemes in Section~\ref{subsec:epid}.
There could be multiple \pa{s} and an individual website can decide which \pa{s} to trust.
If a TEE is provisioned by an unsupported \pa, the website would fall back to using CAPTCHAs.
}

\begin{figure}
	\begin{center}
	\begin{sequencediagram}
		\newinst{tee}{\tee}
		\newinst[5]{pa}{\pa}
			\begin{call}{tee}{get\_group\_private\_key()}{pa}{$G_{sk_{\tee}}$}
				\begin{call}{pa}{request\_attestation()}{tee}{attestation\_report}
				\end{call}
			\end{call}
	\end{sequencediagram}
	\caption{\sys initialization protocol}
	\label{fig:init_phase}
	\end{center}
\end{figure}

\begin{figure*}
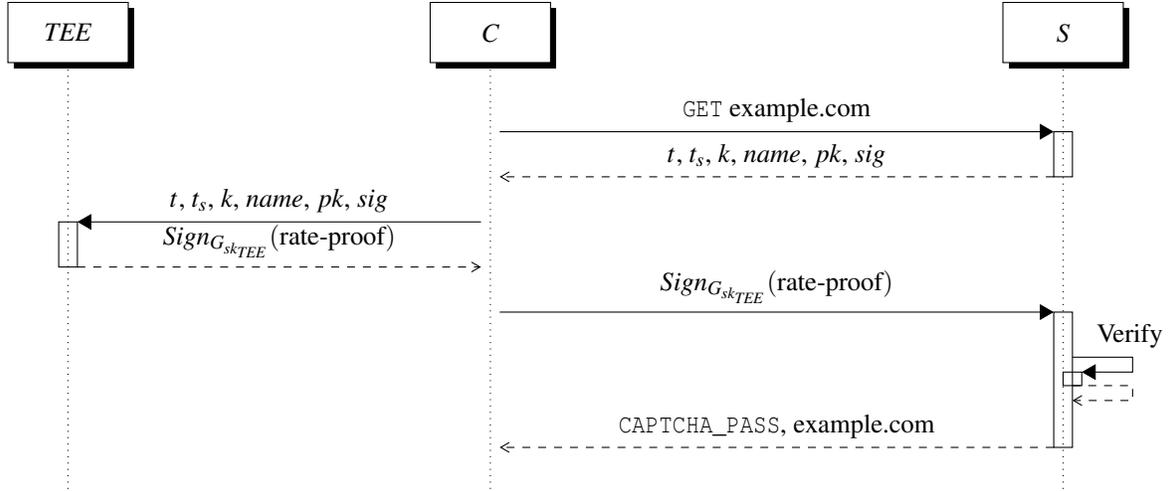

	\begin{center}
	\begin{sequencediagram}
		\newinst{tee}{\tee}
		\newinst[4]{cl}{\cl}
		\newinst[6]{sv}{\sv}
			\begin{call}{cl}{\texttt{GET} example.com}{sv}{$t$, $t_s$, $k$, $name$, $pk$, $sig$}
			\end{call}
			\begin{call}{cl}{$t$, $t_s$, $k$, $name$, $pk$, $sig$}{tee}{$Sign_{G_{sk_{\tee}}}(\text{rate-proof})$}
			\end{call}
			\begin{call}{cl}{$Sign_{G_{sk_{\tee}}}(\text{rate-proof})$}{sv}{\texttt{CAPTCHA\_PASS}, example.com}
			\begin{callself}{sv}{Verify}{}\end{callself}
			\end{call}
	\end{sequencediagram}
	\caption{\sys CAPTCHA-avoidance protocol}
	\label{fig:captcha_avoidance}
	\end{center}
\end{figure*}

Once the TEE is provisioned, the client can begin to use \sys{} when visiting supported websites, as shown in Figure~\ref{fig:captcha_avoidance}.
Specifically, when serving a page, the server includes the following information: a timestamp $t$, a threshold \thr (including start time $t_s$ and count $k$), the name of the list (or \sysglobal for the global list), and (optionally) a public key and signature for rates that enforce a same-origin policy.
The client uses this information to request a rate-proof from their TEE.
If the client's rate is indeed below the threshold, the TEE produces the rate-proof, signed with its group private key.
The client then sends this to the server in lieu of solving a CAPTCHA.

\subsubsection{TEE Design} \label{subsubsec:teedesign}
To realize the conceptual design above, the client's TEE would ideally store all timestamps indefinitely in integrity-protected and rollback-protected memory. However, as discussed above, current TEEs fall short of this idealized representation, since they have limited integrity-protected memory and a limited number of hardware counters for rollback protection.
To overcome this challenge, we store all data outside the TEE, e.g., in a standard database.
To prevent dishonest clients from modifying this data, we use a combination of hash chains and Merkle Hash Trees (MHTs) to achieve integrity and rollback-protection. 

\begin{figure}[t]
	\includegraphics[width=\linewidth]{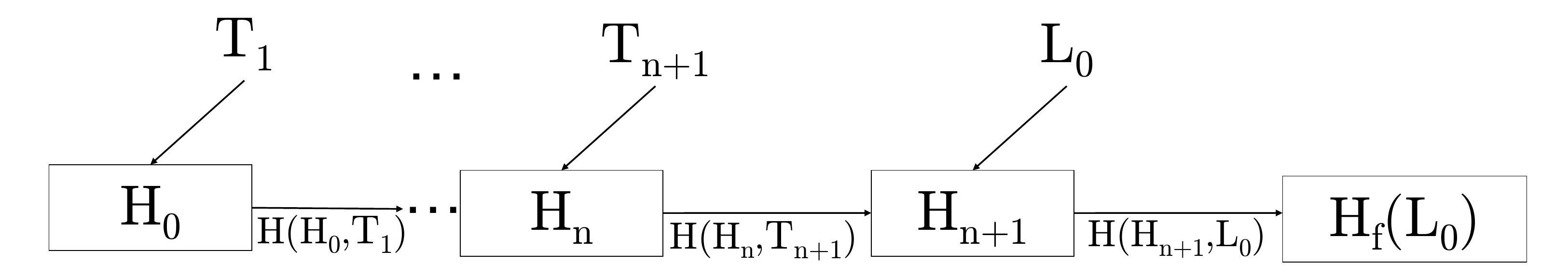}
	\caption{\texttt{final\_hash} calculation from timestamps $T_j$ and list info $L_0$. $H_0 = H(T_0)$.}
	\label{fig:hc}
\end{figure}

\noindent \textbf{Hash chains of timestamps.}
To protect integrity of stored timestamps, we compute a hash chain over each list of timestamps.
We store intermediate value of the hash chain along with each timestamp.
The head of the hash chain is hashed together with the list information (list name and public key) to generate the \texttt{final\_hash} value shown in Figure \ref{fig:hc}.

\begin{figure}[t]
	\includegraphics[width=\linewidth]{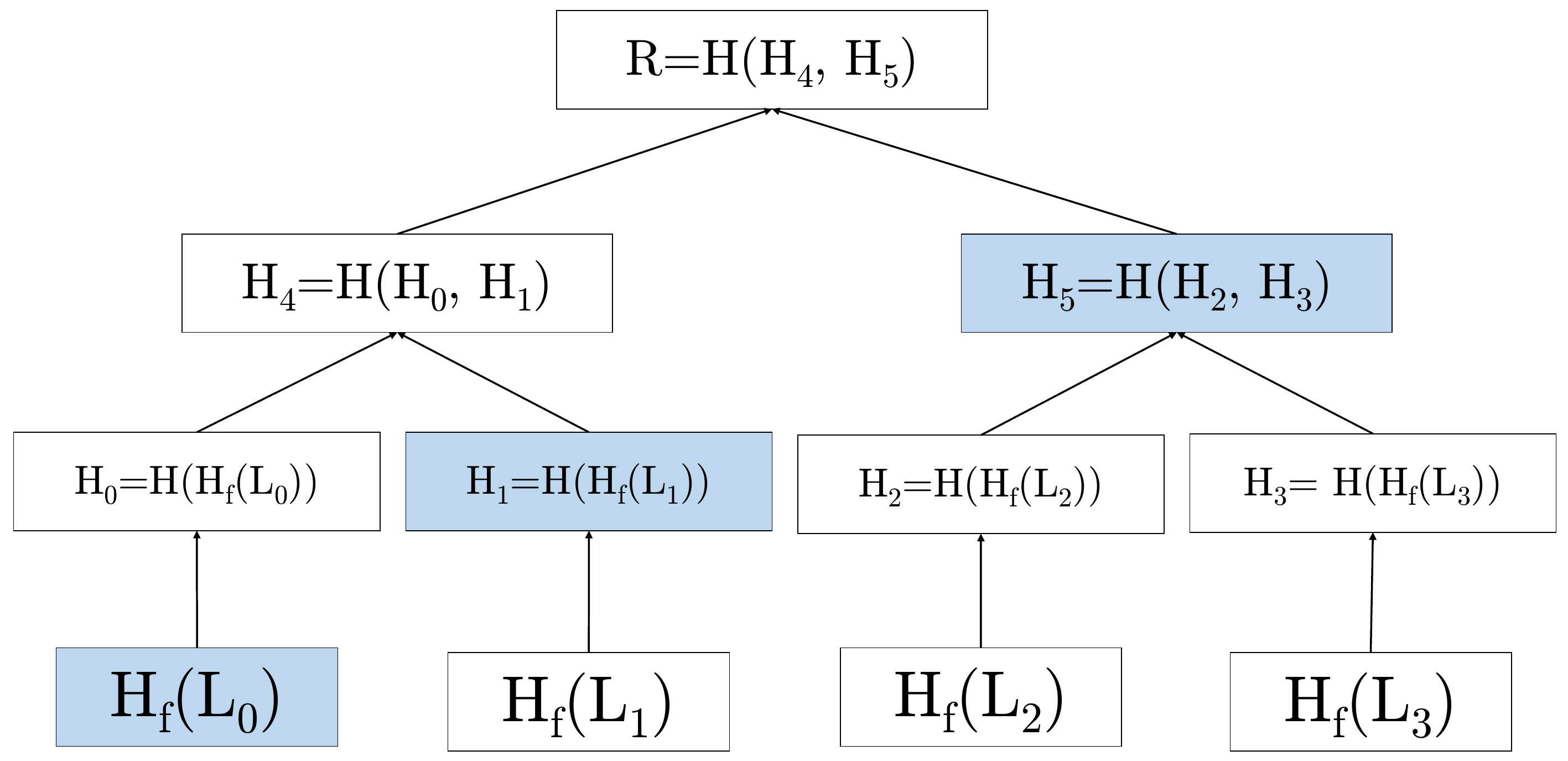}
	\caption{Merkle Tree from timestamp hash chain heads. $H_f({L_i})$ refers to the final hash for the list $L_i$.
	The nodes in blue refer to the inclusion proof path of the final hash for the list $L_0$.}
	\label{fig:mt}
\end{figure}

\noindent \textbf{MHT of lists.}
Thanks to the hash chain scheme, the TEE only needs to provide integrity and rollback-protected storage for the \texttt{final\_hash} values of each list.
It would be possible for the TEE to seal each of these values individually. However, in order to achieve rollback-protection, the TEE would then need a separate hardware monotonic counter for each list, since the lists may be updated independently.
In a real-world deployment, the number of lists is likely to exceed the number of available hardware counters, e.g., 256 per enclave in SGX.
To overcome this challenge, we combine the list into a Merkle Hash Tree (MHT).
Specifically, the \texttt{final\_hash} value of each list becomes a leaf in the MHT, as shown in Figure~\ref{fig:mt}.
With this arrangement, the TEE only needs to provide integrity and rollback-protected storage for the MHT root, which can be easily achieved using sealing and a single hardware monotonic counter.

\subsubsection{Producing a Rate-Proof} \label{subsubsec:rateproof}
The TEE first needs to verify the integrity of its externally-stored data structures (i.e., hash chains and MHT described above), and if successful, update these with the new timestamp and produce the rate-proof, as follows:

\textbf{1. TEE inputs.}
The client supplies its TEE with the all timestamps in the named list that are greater than or equal to the server-defined start time $t_s$.
The client also supplies the largest timestamp that is smaller than $t_s$, which we denote $t_{s-\delta}$, and the intermediate value of the hash chain up to, but not including, $t_{s-\delta}$.
The client supplies the sealed MHT root, and intermediate hashes required to verify that the list's \texttt{final\_hash} value is in the MHT.

\textbf{2. Hash chain checks.}
The TEE first checks that $t_{s-\delta}$ is smaller than $t_s$ and then recomputes the hash chain over included timestamps in order to reach the \texttt{final\_hash} value.
During this process, it counts the number of included timestamps and checks that this is less than the value $k$ specified in the threshold.
The inclusion of one timestamp outside the requested range ($t_{s-\delta}$) ensures that the TEE has seen all timestamps within the range.
This process requires $\mathcal{O}(n)$ hashes, where $n$ is the number of timestamps in the requested range.

\textbf{3. MHT checks.}
The TEE then unseals the MHT root and uses the hardware counter to verify that it is the latest version.
The TEE then checks that the calculated \texttt{final\_hash} value is indeed a leaf in the MHT.
This process requires $\mathcal{O}(log(s))$ hashes, where $s$ is the number of lists.
Including the list name in the \texttt{final\_hash} value ensures that the timestamps have not been substituted from another list.
If the list has an associated public key, the TEE uses this to verify the signature on the server's request.

\textbf{4. Starting a new list.}
\changed{If the rate-proof is requested over a new list (e.g., when the user firsts visits a website), the TEE must also verify that the list name does not appear in any MHT leaves.}
In this case, the client supplies the TEE with all list names and their corresponding \texttt{final\_hash} values.
The TEE reconstructs the full MHT and checks that the new list name does not appear.
This requires $\mathcal{O}(s)$ string comparisons and hashes.

\textbf{5. Updating a list.}
If the above verification steps are successful, the TEE checks that the new timestamp $t$ supplied by the server exceeds the latest timestamp in the specified list.
If so, the TEE adds $t$ to the list, computes the new \texttt{final\_hash} value and the new MHT root.
The new root is sealed alongside the TEE's group private key.
The TEE then produces a signed rate-proof, using its group private key.
The rate-proof includes a hash of the original request provided by the server, thus confirming that the TEE checked the rate and added the server-supplied timestamp.
The TEE returns the rate-proof to the client, along with the new sealed MHT root for the client to store.
In the above design, the whole process of producing the rate-proof can be performed in a single call to the TEE, thus minimizing the overhead of entering/exiting the TEE.

\changed{
\subsubsection{Reducing Client-Side Storage} \label{subsubsec:prune}
The number of timestamps stored by \sys{} grows as the client visits more websites.
However, in most use-cases, it is unlikely that the server will request rate-proofs going back beyond a certain point in time $t_P$.

To reduce client-side storage requirements, we provide a mechanism to \emph{prune} a client's timestamp list by merging all timestamps prior to $t_P$.  Specifically, the server can include $t_P$ in any rate-proof request, and upon receiving this, the client's TEE counts and records how many timestamps are older than $t_P$. 
The old timestamps and associated intermediate hash values can then be deleted from the database. 
In other words, the system merges all timestamps prior to $t_P$ into a single count value $c_P$.
The TEE stores $t_P$ and the count value in the database outside the TEE and protects their integrity by including both values in the list information that forms the MHT leaf.
Pruning can be done repeatedly: when a new pruning request is received for $t_{P'} > t_P$, \sys fetches and verifies all timestamps up to $t_{P'}$ and adds these to $c_{P}$ to create $c_{P'}$.
It then replaces $t_P$ and $c_{P}$ with $t_{P'}$ and $c_{P'}$ respectively.

This pruning mechanism does not reduce security of \sys{}. If the server does request a rate-proof going back beyond $t_P$, \sys{} will include the full count of timestamps stored alongside $t_P$.
This is always greater than or equal to the actual number of timestamps; thus, there is no incentive for the server to abuse the pruning mechanism. Similarly, even if a malicious client could trigger this pruning (i.e., assuming the list is not associated to the server's public key), there is no incentive to do so because it would never decrease the number of timestamps included in rate-proofs.

Since the global list \sysglobal{} is used by all websites, the client is always allowed to prune this list to reduce storage requirements.
\sys blocks servers from pruning \sysglobal since this can be used as an attack vector to inflate the client rate by compressing all rates into one value -- thus preventing use of \sys on websites that utilize \sysglobal.
Thus, we expect pruning of \sysglobal to be done automatically by the \sys host application or browser extension.
}

\section{Implementation} \label{sec:impl}
\begin{figure}
	\includegraphics[width=.47\textwidth]{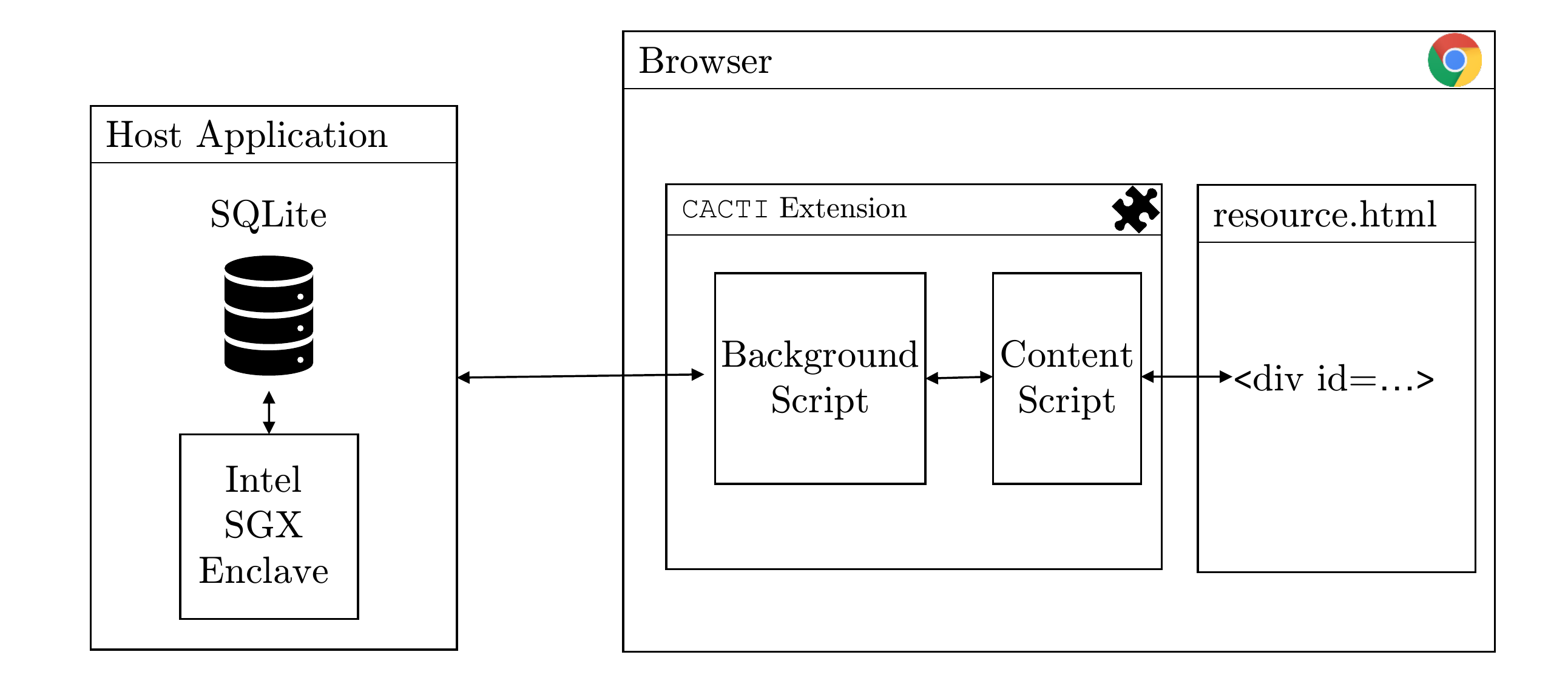}
	\caption{Overview of \sys client-side components.}
	\label{fig:overview}
\end{figure}

We now describe the implementation of the \sys design presented in the previous section.
We focus on proof-of-concept implementations of: client-side browser extension, native host application, and 
\sys{} TEE, as shown in Figure~\ref{fig:overview}. 
\changed{Finally, we discuss how \sys is integrated into websites.}

\subsection{Browser Extension}\label{subsec:extension}
The browser extension serves as a bridge between the web server and our host application.
We implemented a proof-of-concept browser extension for the Chrome browser (build 79.0.3945.130)~\cite{chrome}.
Chrome extensions consist of two parts: a content script and a background script.
\begin{itemize}
	\item \textbf{Content script:}
	\changed{scans} the visited web page for an HTML \texttt{div} element with the id \texttt{\sys-div}.
	If the page contains this, the content script parses the parameters it contains and sends them to the background script.
	\item \textbf{Background script:}
	we use Chrome Native Messaging to launch the host application binary when the browser is started and maintain an open port~\cite{port} to the host application until the browser is closed.
	The background script facilitates communication between the content script and the host application.
\end{itemize}

\changed{
\noindent\textbf{User notification.}
The browser extension is also responsible for notifying the user about requests to access \sys{}.
Notifications can include information, such as server's domain name, timestamp to be inserted, and threshold used to generate the rate-proof.
By default, the background script notifies the user whenever a server requests to use \sys{}, and waits for user confirmation before proceeding.
This prevents malicious websites from abusing \sys{} by adding multiple timestamps without user permission (for possible attacks, see Section \ref{subsec:seceval}). 
However, asking for user confirmation for every request could cause UI fatigue.
Therefore, \sys{} could allow the user to choose from the following options: 
(1) \emph{Always ask} (the default),
(2) \emph{Ask only upon first visit to site},
(3) \emph{Only ask for untrusted sites}, 
(4) \emph{Only ask for more than $x$ requests per site per time period},
and (5) \emph{Never ask}.
Advanced users can also modify our extension or code their own extension to enforce arbitrary policies for requesting user confirmation.
The notification is displayed using Chrome's Notification API~\cite{chromenotify}.
}

\subsection{Host Application}\label{subsec:ha}
The host application running on the client is responsible \changed{for}: 
(1) creating the \sys TEE, which we implement as an SGX enclave, and exposing its \ecall API to the browser extension; 
(2) storing (and forwarding) timestamps and additional integrity information for secure calculation of rate-proofs (to the enclave); 
and (3) returning the enclave's output to the browser extension.

The host application is implemented in C and uses Chrome Native Messaging~\cite{nativeMessaging} to communicate with the browser extension.
Since Chrome Native Messaging only supports communication with JSON objects, the host application uses a JSON parser to extract parameters to the API calls.
We used the JSMN JSON parser~\cite{jsmn}.
Moreover, the host application implements the Chrome Native Messaging protocol~\cite{nmprotocol} and communicates with the browser extension using Standard I/O (\texttt{stdio}), since this is currently the only means to communicate between browser extensions and native applications.

The host application stores information in an SQLite database.
This database has two tables: \texttt{LISTS} stores the list names and associated public keys, and \texttt{TIMESTAMPS} stores all timestamps and intermediate values of the hash chains. 
For each rate-proof request, the host application queries the database and provides the data to the enclave.

Since the timestamps are stored unencrypted, we use existing features of the SQLite database to retrieve only the necessary range of timestamps for a given list. 
Note that since data integrity is maintained through other mechanisms (i.e., hash chains and MHT), the mechanism used by the host application to store this data does not affect the security of the system.
Alternative implementations could use different database types and/or other data storage approaches.
Instead of hash chains and MHTs, it is possible to use a database managed by the enclave, e.g., EnclaveDB~\cite{enclavedb}.
However, this would increase the amount of code running inside the enclave, thus bloating the trusted code base (TCB).

\subsection{SGX Enclave} \label{subsec:enc}
We implemented the TEE as an SGX enclave using the OpenEnclave SDK~\cite{oe} v0.7.0.
OpenEnclave was selected since it aims to unify the programming model across different types of TEEs.
The process of requesting a rate-proof is implemented as a single \texttt{get\_rate} \texttt{ECALL}.
For timestamps, we use the UNIX time which denotes the number of seconds elapsed since the UNIX Epoch (midnight 1/1/1970) and is represented as a 4-byte signed integer.
We use cryptographic functions from the {\tt mbed} TLS library~\cite{mbedtlsgithub} included in OpenEnclave.
Specifically, we use SHA-256 for all hashes and ECDSA for all digital signatures. For EPID signatures, we use Intel EPID SDK (v7.0.1)~\cite{epidgithub} \changed{with the performance-optimized version of Intel Integrated Performance Primitives (IPP) Cryptography library \cite{ipp-optimized}}.
We use a formally-verified and platform-optimized MHT implementation from EverCrypt~\cite{cryptoeprint:2019:757}.
As an optimization, if the MHT is sufficiently small, we can cache fully inside the enclave.
When a request for a rate-proof is received, the enclave recalculates the timestamp hash chain and then directly compares the \texttt{final\_hash} value with the corresponding leaf in the cached MHT.

One limitation of OpenEnclave is its lack of support for SGX hardware monotonic counters.
Therefore, we cannot include these in this proof-of-concept implementation.
However, a production implementation can easily include hardware counter functionality.
\changed{
Although our implementation uses SGX, \sys{} can be realized on any suitable TEE.
For example, OpenEnclave is currently being updated to support ARM TrustZone.
When this version is released, we plan to port the current implementation to TrustZone, with minimal expected modifications.
}

\changed{
\subsection{Website Integration} \label{subsec:integration}
Integrating \sys into a website involves two aspects: sending the rate-proof request to the client, and verifying the response.
The server generates the rate-proof request (see Section \ref{subsubsec:comm}) and encodes it as \texttt{data-*} 
attributes in the \texttt{CACTI-div} HTML \texttt{div}.
The server also includes the URL to which the generated rate-proofs should be sent.
The browser extension determines whether the website supports \sys by looking for the \texttt{CACTI-div} element.
The server implements an HTTP endpoint for receiving and verifying rate-proofs . 
If the verification succeeds, this endpoint notifies the website and the user is granted access.

Integrating \sys into a website is thus very similar to using existing CAPTCHA systems.
For example, reCAPTCHA adds the \texttt{g-recaptcha} HTML \texttt{div} to the page, and implements various 
endpoints for receiving and verifying the responses~\cite{recaptchadev}.
We evaluate server-side overhead of \sys, in terms of both processing and data transfer requirements, in Section~\ref{sec:eval}.
}

\section{Evaluation} \label{sec:eval}
We now present and discuss the evaluation of \sys.
We start with a security analysis, based on the threat model and requirements defined in Section~\ref{sec:tmodreq}.
Next, we evaluate performance of \sys in terms of \changed{latency and bandwidth.}
Finally, we discuss \sys deployability issues. 

\subsection{Security Evaluation} \label{subsec:seceval}
\textbf{Data integrity \& rollback attacks.}
Since timestamps are stored outside the enclave, a malicious host application can try to modify this data, or roll it back to an earlier version.
If successful, this might trick the enclave into producing falsified rate-proofs.
However, if any timestamp is modified outside the enclave, this would be detected because the final hash value of the hash chain would not match the corresponding MHT leaf.
Assuming a suitable collision-resistant cryptographic hash function, it is infeasible for the malicious host to find alternative hash values matching the MHT root. 
Similarly, a rollback attack against the MHT is detected by comparing the included counter with the hardware monotonic counter.

\textbf{Timestamp omission attacks.}
A malicious application can try to provide the enclave with only a subset of the timestamps for a given request, e.g., to pretend to be below the threshold rate. Specifically, the host could try to omit one or more timestamps at the start, in the middle, and/or at the end, of the range.
If timestamps are omitted at the start, the enclave detects this when it checks that the first timestamp supplied by the host is \emph{prior to} the start time of request $t_s$.
If timestamps are omitted in the middle (or at the end) of the range, the final hash value will not match the value in the MHT leaf.

\textbf{List substitution attacks.}
A malicious client might attempt to use a timestamp hash chain from a different list, or claim that the requested list does not exist.
The former is prevented by including list information (list name and public key) in the final hash calculation.
If there is a mismatch between the name and the timestamp chain, the resulting final hash would not exist in the MHT.
For the latter, when the host calls the enclave's \texttt{get\_rate} function for a new list, the enclave checks the names of all lists in the MHT to ensure that the new list name does not already exist.

\textbf{TEE reset attacks.}
A malicious client might attempt to delete all stored data, including the sealed MHT root, in order to reset the TEE.
Since the group private key received from the provisioning authority is sealed together with the MHT root, it is impossible to delete one and not the other.
Deleting the group private key would force the TEE to be re-provisioned by the provisioning authority, which may apply its own rate-limiting policies on how often a given client can be re-provisioned.

\changed{\textbf{\sys{} Farms.} Similar to CAPTCHA farms, a multitude of devices with TEE capabilities could be employed to satisfy rate thresholds set by servers.
However, this would be infeasible because: \textit{(1)} \sys enclaves would stop producing rate-proofs after reaching server thresholds and would thus require a TEE reset and \sys{} re-provisioning -- which is a natural rate limit; \textit{(2)} the cost of purchasing a device would be significantly higher than CAPTCHA solving costs.
For example, in 2010 it cost \$1 to have $1,000$ CAPTCHAs solved by a CAPTCHA farm~\cite{motoyama2010re} -- currently the cheapest service charges \$1.8 for solving $1,000$ reCAPTCHAs \cite{antiCaptcha}\footnote{See a comparison of CAPTCHA solving services \cite{captchaSolverComp}}, while a low-end bare-bones CPU with SGX support alone costs $\approx\$70$~\cite{intelProcessor}.
}

\textbf{Client-side malware.}
A more subtle variant of the reset attack can occur if malware on the client's own system corrupts or deletes TEE data.
This is a type of denial-of-service (DoS) attack against the client.
However, defending against such DoS attacks is beyond the scope of this work, since this type of malware would have many other avenues for causing DoS, e.g., deleting critical files.

\textbf{Other DoS attacks.}
A malicious server might try to mount a DoS attack against an unsuspecting client by inserting a timestamp for a future time.
If successful, the client would be unable to insert new timestamps and create rate-proofs for any other servers, since the enclave would reject these timestamps as being in the past.
This attack can be mitigated if the client's browser extension and/or host application simply check that the server-provided timestamp is not in the future.

\changed{
\textbf{Client tracking.}
A malicious server (or group of servers) might attempt to track clients by sending multiple requests for rate-proofs with different thresholds in order to learn the precise number of timestamps stored by the client.
A successful attack of this type could potentially reduce the client's anonymity set to only those clients with the same rate.
However, this attack is easy to detect by monitoring the thresholds sent by the server.
A more complicated attack targeting a specific client is to send an excessive number of successful rate-proof requests in order to increase the client's rate.
The goal is to reduce the size of the target's anonymity set.
This attack is also easy to detect or prevent by simply rate-limiting the number of increments accepted from a particular server.
Note that the window of opportunity for this targeted attack is limited to a single session, because malicious servers cannot reliably re-identify the user across multiple sessions (since this is what the attack is trying to achieve).
The above attacks cannot be improved even if multiple servers collude.
}

\changed{
\textbf{Rogue \pa{s}.} 
A malicious \pa{} might try to compromise or diminish client privacy. 
However, this is prevented by \sys's use of the EPID protocol~\cite{epid}.
Specifically, due to the BBS+ signature scheme~\cite{au2006constant} during EPID key issuance, clients' private keys are never revealed to \pa{s}. 
Also, EPID's signature-based revocation mechanism does not require member private keys to be revealed. 
Instead, signers generate zero-knowledge proofs showing that they are not on the revocation list.
Therefore, client privacy does not depend on any \pa business practices, e.g., log deletion or identifier blinding. 

Each website has full discretion to decide which \pa{s} it trusts; if a server does not trust the PA who issued the member private key to the
TEE, it can simply fall back to CAPTCHAs. 
This provides no advantage to attackers, and websites can be as conservative as they like. 
If higher levels of assurance are required, \pa{s} can execute within TEEs and provide attestation of correct behavior; 
we defer the implementation of this optional feature to future work.
}

\changed{Overall, we claim that \sys{} meets all security requirements defined in Section~\ref{sec:tmodreq} and significantly increases
the adversary's cost to perform DoS attacks}.
Specifically, the \textbf{Unforgeability} requirement is satisfied since it is impossible for the host to perform rollback, timestamp exclusion and list substitution attacks.
\textbf{Client privacy} is achieved because the rate-proof does not reveal the actual number of timestamps included, and is signed using a group signature scheme.

\subsection{Latency Evaluation} \label{subsec:le}
We conducted all latency experiments on an Intel NUC Kit NUC7PJYH~\cite{nuc} with an Intel Pentium Silver J5005 Processor (4M Cache, up to 2.80 GHz); 4 GB DDR4-2400 1.2V SO-DIMM Memory; running Ubuntu 16.04 with the Linux 4.15.0-76-generic kernel Intel SGX DCAP Linux 1.4 drivers.

Recall that the host application is responsible for initializing the enclave, fetching data necessary for enclave functionality, performing \ecall{s}, and finally updating states according to enclave output.
Therefore, we consider the latency in the following four key phases in the host application:
\begin{compactitem}
	\item \textit{Init-Enclave}: Host retrieves the appropriate data from the database and calls \texttt{init\_mt} \ecall that initializes the
	MHT within the enclave.\footnote{Init-Enclave is done only when the enclave starts.}
	\item \textit{Pre-Enclave}: Host retrieves the required hashes and timestamps from the database.
	\item \textit{In-Enclave}: Host calls the \texttt{get\_rate} \ecall. This phase concludes when the \ecall returns.
	\item \textit{Post-Enclave}: Host updates/inserts the data it received from the enclave into the database.
\end{compactitem}
We investigated the latency impact by varying (1) the number of timestamps in the rate-proof (Section~\ref{lvst}), and (2) the number of lists in the database (Section~\ref{subsubsec:lvss}).
We evaluated the end-to-end latency in Section~\ref{subsubsec:overall}.
Unless otherwise specified, each measurement is the average of 10 runs.

\changed{
\noindent\textbf{Note:} The ECDSA and EPID signature operations are, by far, the dominant contributors to latency.
However, they represent a fixed latency overhead that does not vary with the number of timestamps or servers.
Therefore, for clarity's sake, figures in the following sections do not include these operations.
We analyze them separately in Section~\ref{subsubsec:cryptoeval}.
}

\subsubsection{Varying Number of Timestamps in Query} \label{lvst}
\begin{figure*}[t]
\centering
\noindent
\begin{minipage}{0.49\textwidth}
	\includegraphics[width=\columnwidth]{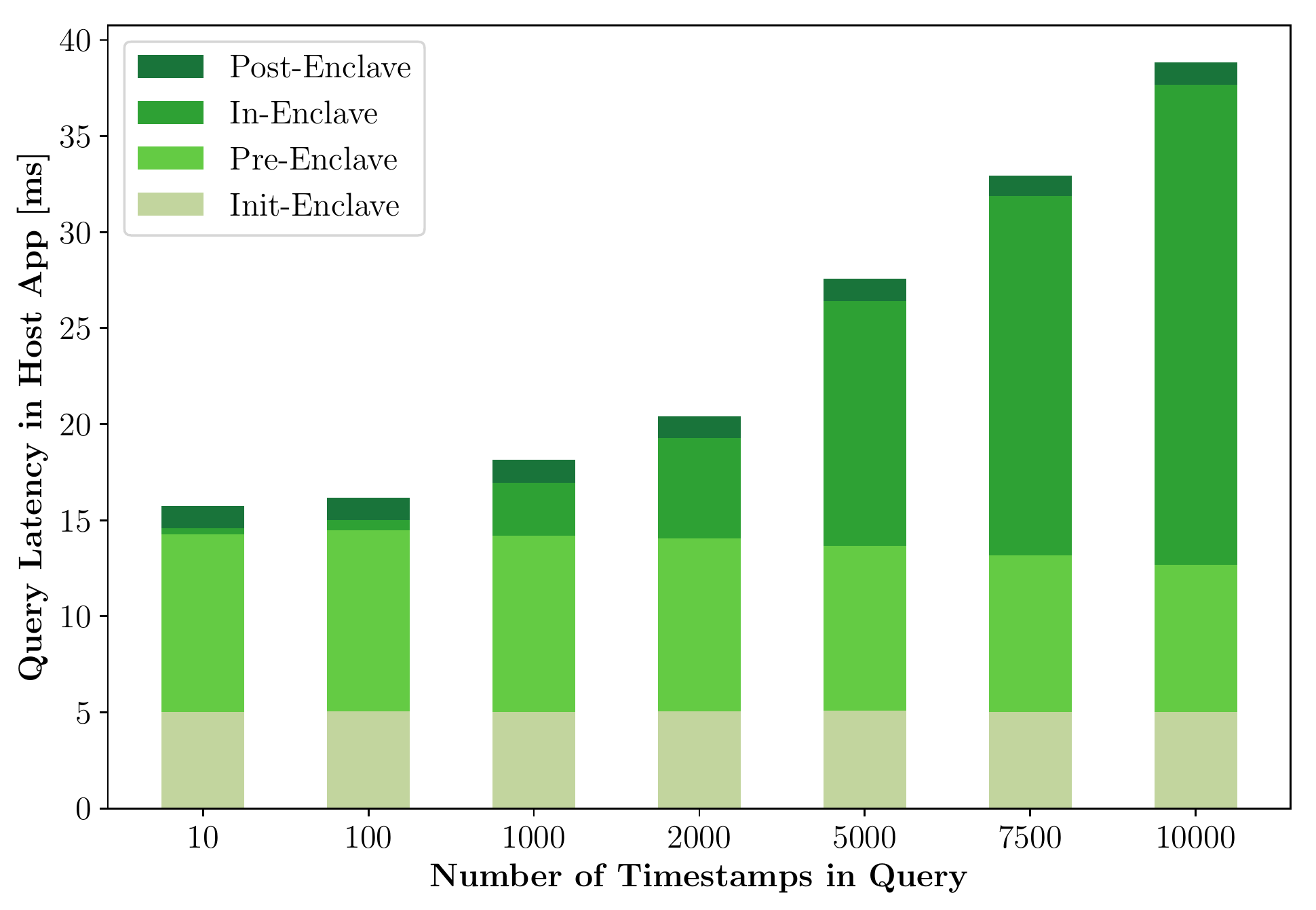}
	\caption{\changed{Query latency for different numbers of timestamps.}}
	\label{fig:lvst}
\end{minipage}\hfill
\begin{minipage}{0.49\textwidth}
	\includegraphics[width=\columnwidth]{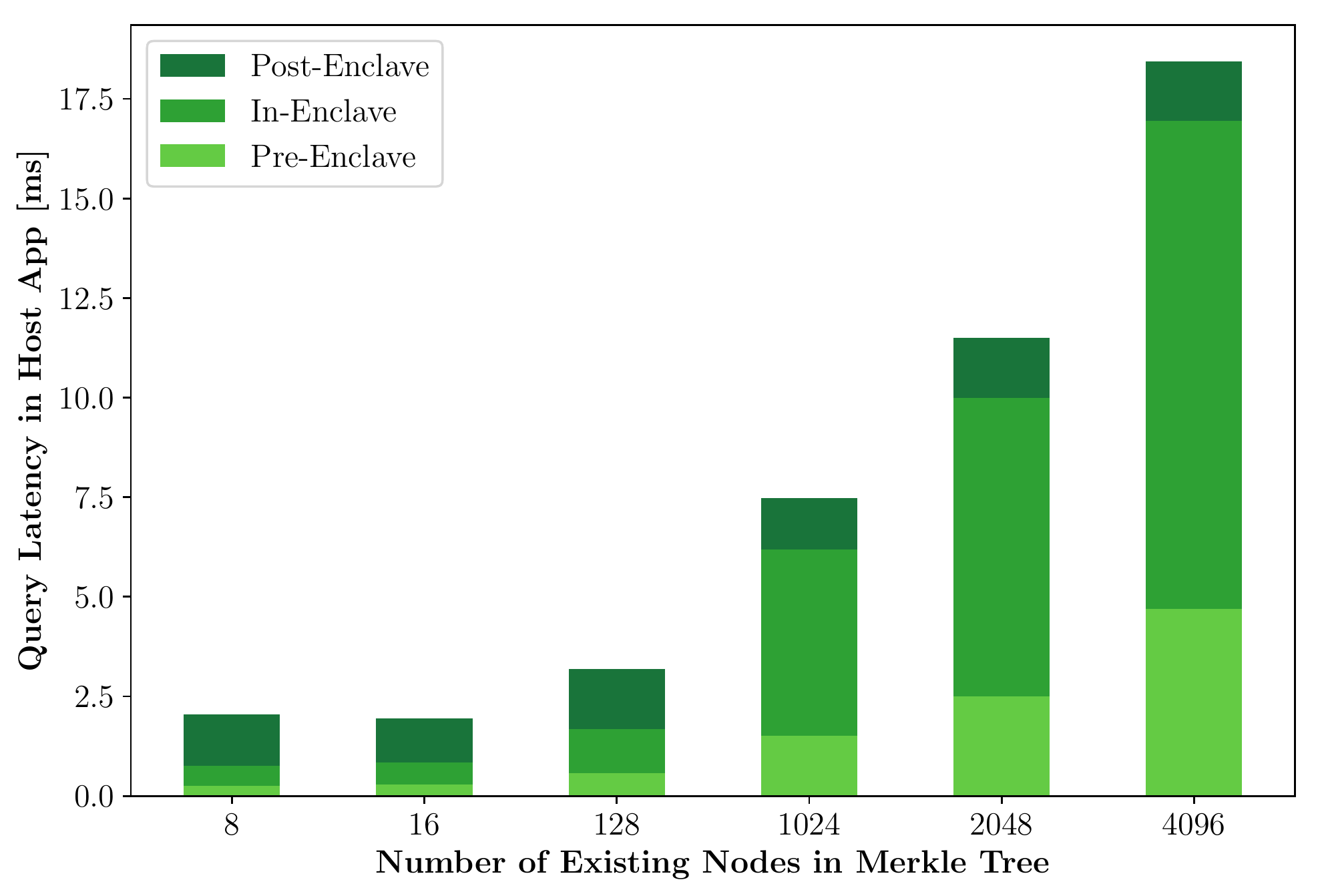}
	\caption{\changed{Query latency when adding a new list.}}
	\label{fig:mtpopulate}
\end{minipage}

\vspace{3mm}
\noindent
\begin{minipage}{0.49\textwidth}
	\includegraphics[width=\columnwidth]{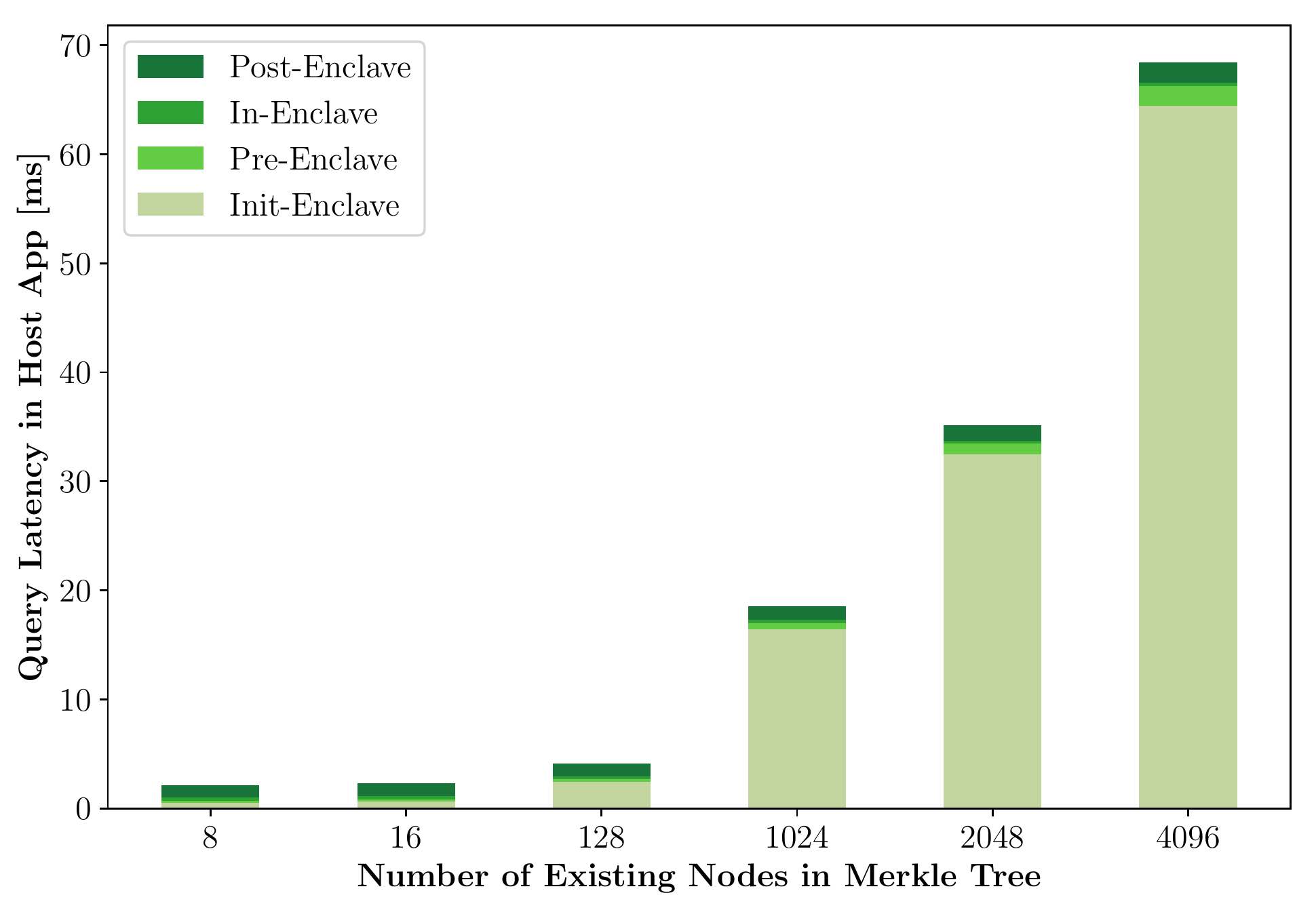}
	\caption{\changed{Query latency for updating an existing list.}}
	\label{fig:lvsn_with_init}
\end{minipage}\hfill
\begin{minipage}{0.49\textwidth}
	\includegraphics[width=\columnwidth]{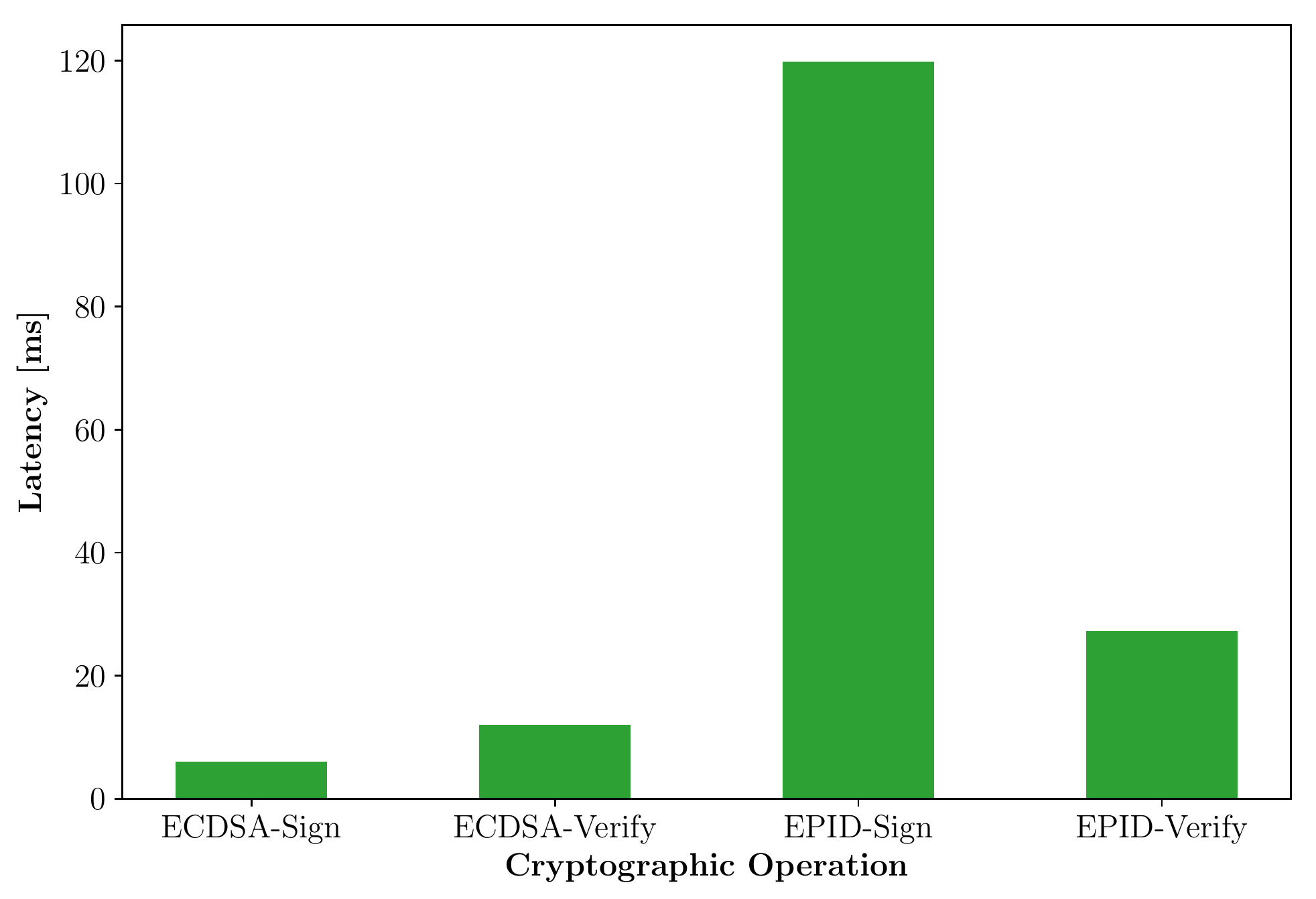}
	\caption{\changed{Latency of signature operations.}}
	\label{fig:crypto_eval}
\end{minipage}
\end{figure*}

We measured the effect of varying the number of timestamps included in the query, while holding the number of lists constant.
As shown in Figure~\ref{fig:lvst}, query latency increases linearly with the number of timestamps included in the query.
The most notable increase is in the pre-enclave phase, since this involves retrieving the requested timestamps from the database.
The in-enclave phase also increases slightly since the enclave must calculate a longer hash chain.
However, even with 10,000 timestamps in a query, the total latency only reaches \textasciitilde40~milliseconds (excluding signature operations).

\subsubsection{Varying Number of Lists} \label{subsubsec:lvss}
Next, we varied the number of lists while holding the number of timestamps fixed at one per list.
We considered two separate scenarios: adding a new list and updating an existing list.

\textbf{Adding a new list.}
As shown in Figure~\ref{fig:mtpopulate}, the latency for the pre-enclave phase is very low because we optimize the host to skip the expensive \texttt{TIMESTAMPS} table look up operation if the host knows that this is a new list.
The in-enclave phase increases as the number of lists increases due to the string comparison operations performed by the enclave to prevent list substitution attacks.
However, this phase can be optimized by sorting the server names inside the enclave during initial MHT construction.
The post-enclave latency is due to the cost of adding entries to the \texttt{TIMESTAMPS} table.

\textbf{Updating an existing list.}
As shown in Figure~\ref{fig:lvsn_with_init}, the latency of the init-enclave phase increases as the number of lists increases.
This is expected, since the enclave reconstructs the MHT in this phase.
The pre-enclave phase also increases slightly due to the database operations.

\subsubsection{Signature Operation Latency} \label{subsubsec:cryptoeval}
Evaluation results presented thus far have not included the ECDSA signature verification or EPID signature creation operations.
Specifically, the server creates an ECDSA signature on the request, which the enclave verifies.
The enclave creates an EPID group signature on the response, which the server verifies using the EPID group public key.
The average latencies over 10 measurements for these four signature operations are shown in Figure~\ref{fig:crypto_eval}.
\changed{We can see that the EPID group signature generation operation is an order magnitude slower compared to the other cryptographic operations including EPID group signature verification.}
The latency of our enclave is thus dominated by the EPID signature generation operation.

\begin{table*}[t]
	\centering
	\caption{End-to-End Latency of \sys for different numbers of timestamps and lists. The \emph{Browser} column represents the latency of the browser extension marshalling data to and from the host application. The other columns are as described above.} \label{tab:latency}
	\newcolumntype{R}{>{\raggedleft\arraybackslash}X}
	\begin{tabularx}{\textwidth}{p{26mm} r r r r r r R}
	\toprule
	& \textbf{ECDSA-Sign} & \textbf{Browser} & \textbf{Pre-Enclave} & \textbf{In-Enclave} & \textbf{Post-Enclave} & \textbf{EPID-Verify} & \textbf{Total}  \\ \midrule
	10,000 timestamps in 1 list & 6.3~ms & 15.2~ms & 7.7~ms & \changed{181.7~ms} & 1.0~ms & \changed{27.3~ms} & \changed{239.2~ms} \\
	4,096 lists with 1 timestamp each  & 6.3~ms & 15.2~ms & 1.8~ms & \changed{157.4~ms} & 2.0~ms & \changed{27.3~ms} & \changed{210.0~ms} \\ \bottomrule
	\end{tabularx}
\end{table*}

\subsubsection{End-to-End Latency} \label{subsubsec:overall}
Table~\ref{tab:latency} shows the end-to-end latency (excluding network communication) from when the server begins generating a request until it has received and verified the response from the client.
In both settings, the end-to-end latency is below \changed{250~milliseconds}.
The latency will be lower if there are fewer lists or included timestamps.
Compared to other types of CAPTCHAs, image-based CAPTCHAs take \textasciitilde10 seconds to solve~\cite{bursztein2010good} and behavior-based reCAPTCHA takes \textasciitilde400~milliseconds, although this might change depending on the client's network latency.

\subsection{Bandwidth Evaluation} \label{subsec:be}
\changed{
We measured the amount of additional data transferred over the network by different types of CAPTCHA techniques.
Minimizing data transfer is critical for both servers and clients.
We compared \sys against image-based and behavior-based reCAPTCHA~\cite{recaptcha} (see Figure~\ref{fig:captchas}).
The former asks clients (one or more times) to find and mark certain objects in a given image or images, whilst the latter requires clients to click a button.
To isolate the data used by reCAPTCHA, we hosted a webpage with the minimal auto-rendering reCAPTCHA example~\cite{recaptchadev}.
We visited this webpage and recorded the traffic using the Chrome browser's debugging console.
}

\changed{
Table~\ref{tab:bw} shows the additional data received and sent by the client to support each type of CAPTCHA.
Image-based reCAPTCHA incurs the highest bandwidth overhead since it has to download images, often multiple times.
Although not evaluated here, text-based CAPTCHAs also use images and would thus have a similar bandwidth overhead.
Behavior-based reCAPTCHA downloads several client-side scripts.
Both types of reCAPTCHA made several additional connections to Google servers.
Overall, \sys achieves at least a 97\% reduction in client bandwidth overhead compared to contemporary reCAPTCHA solutions.
}
\begin{table}[t]
	\caption{\changed{Additional data received and sent by the client for image-based and behavior-based reCAPTCHA, compared with \sys.}} \label{tab:bw}
	\centering
	\newcolumntype{R}{>{\raggedleft\arraybackslash}X}
	\changed{\begin{tabularx}{\columnwidth}{l R R R}
		\toprule
						& \textbf{Received} & \textbf{Sent} & \textbf{Total} \\ \midrule
		Image-based     & 140.05~kB  & 28.97~kB   & 169.02~kB \\ 
		Behavior-based  &  54.38~kB  & 26.12~kB   &  80.50~kB \\ 
		\sys            &   0.82~kB  &  1.10~kB   &   1.92~kB \\ \bottomrule
	\end{tabularx}}
\end{table}

\changed{
\subsection{Server Load Evaluation} \label{subsec:sle}
We analyzed the additional load imposed on the server by \sys.
Unfortunately, CAPTCHAs offered as services, such as reCAPTCHA~\cite{recaptcha} and hCAPTCHA~\cite{hCaptcha}, do not disclose their source code and we have no reliable way of estimating their server-side overhead.
Therefore, we compared \sys{} against two open-source CAPTCHA projects published on GitHub (both have more than 1,000 stars and been forked more than a hundred times):

\textbf{dchest/captcha}~\cite{dchestCaptcha} (Figure~\ref{fig:dchest_captcha}) generates image-based text recognition CAPTCHAs consisting of transformed digits with noise in the form of parabolic lines and additional clusters of points.
It can also generate audio CAPTCHAs, which are pronunciations of digits with randomized speed and pitch and randomly-generated background noise.

\textbf{produck/svg-captcha}~\cite{svgCaptcha} (Figure~\ref{fig:svg_captcha}) generates similar image-based text recognition CAPTCHAs, as well as challenge-based CAPTCHAs consisting of simple algebraic operations on random integers.
Noise is introduced by varying the text color and adding parabolic lines.

\begin{figure}
	\begin{subfigure}[b]{.5\textwidth}
		\centering
		\includegraphics[height=1cm]{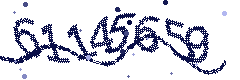}
		\caption{\changed{dchest/captcha image-based CAPTCHA~\cite{dchestCaptcha}.}}
		\label{fig:dchest_captcha}
	\end{subfigure}
	\begin{subfigure}[b]{.5\textwidth}
		\centering
		\includegraphics[height=1cm]{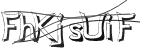}
		\includegraphics[height=1cm]{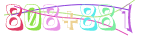}	
		\caption{\changed{produck/svg-captcha image-based CAPTCHAs~\cite{svgCaptcha}.}}
		\label{fig:svg_captcha}
	\end{subfigure}
	\caption{\changed{CAPTCHAs generated using open-source libraries.}}
	\label{fig:sleCAPTCHAs}
\end{figure}

Table~\ref{tab:sle} shows the time to generate different types of CAPTCHAs using the above libraries with typical configuration parameters (e.g., eight characters for text CAPTCHAs).
Since CAPTCHA verification with these libraries is a simple string comparison, we assume this is negligible.
\sys{'s} server-side processing is due almost entirely to the EPID signature verification operation.
We expect that this time could be improved by using more optimized implementations of this cryptographic operation.
Additionally, \sys uses significantly less communication bandwidth than other approaches, which also reduces the server load (which is not captured in this measurement).
Most importantly, the biggest gain of \sys is on the user side; saving more than \textasciitilde10 seconds per CAPTCHA for users.

\begin{table}[t]
	\centering
	\caption{\changed{Server-side processing time for generating a CAPTCHA and verifying the response.}}
	\changed{\begin{tabularx}{\columnwidth}{l X r}
	\toprule
	\textbf{Library} & \textbf{Type} & \textbf{Time} \\ \midrule
	\multirow{2}{*}{dchest/captcha}   & Audio  & 13.3 ms \\ 
	                                  & Image-based text & 1.7 ms \\ \midrule
	\multirow{2}{*}{produck/svg-captcha} & Image-based text & 2.2 ms \\
	                                     & Image-based math & 1.4 ms \\ \midrule
	\sys & Rate-proof & 33.6 ms \\ \bottomrule
	\end{tabularx}}
	\label{tab:sle}
\end{table}
}

\subsection{Deployability Analysis} \label{subsec:deployability}
We analyze deployability of  \sys{} by considering changes required from both the server's and client's perspectives:

\textbf{Server's perspective.}
The server will have to make the following changes: 
(1) create and maintain a new public/private key pair and obtain a certificate for the public key, 
(2) add an additional \texttt{div} to pages for which they wish to enable \sys, 
(3) create and sign requests using the private key, and
(4) add an HTTP endpoint to receive and verify EPID signatures.
The server-side deployment could be further simplified by providing the request generation and signature operations as an integrated library. 

\textbf{Client's perspective.}
The client will have to make the following changes:
(1) download and install the \sys native software, and 
(2) download and install the browser extension.
Although \sys requires the client to have a suitable TEE, this is a realistic assumption given the large and increasing deployed base of devices with e.g., ARM TrustZone or Intel SGX TEEs.

\section{Discussion} \label{sec:disc}
\subsection{\pa Considerations} \label{subsec:paconsiderations}
As discussed in Section~\ref{subsec:realizing}, \sys's use of a provisioning authority (\pa) provides the basis for client privacy.
\sys does not prescribe \pa policies.
For example, \pa has the choice of making the initialization phase (Figure~\ref{fig:init_phase}) a one-off operation (e.g., when installing \sys) or a more regular process, depending on its risk appetite.
For instance, if there are attacks or exploits threatening the Intel SGX ecosystem (and consequently security of group member keys), \pa can revoke all group member keys in the group.
This mandates ``on-demand'' re-registration of all enclaves with \pa.
A similar scenario applies if key-rolling is implemented on \pa, i.e., master secret held owned by \pa is updated frequently.
This forces all enclaves to regularly contact \pa and obtain the new group member key.
Frequent key-rolling introduces heavier burden to the clients (although this can be automated), but provides better security.

\subsection{EPID} \label{subsec:epid}
Even though \sys uses EPID group signatures to protect client privacy, \changed{\sys{}} is agnostic to the choice of the underlying signature scheme as long as it provides \changed{signer unlinkability and  anonymity.
}
We also considered other schemes, such as the Direct Anonymous Attestation (DAA)~\cite{brickell2004direct}, which was is used in the Trusted Platform Module (TPM).
However, DAA is susceptible to various attacks~\cite{leung2008possible, rudolph2007covert, brickell2012static} and, due to its design targeting low-end devices, suffers from performance problems.
In contrast, EPID is used in the current Intel remote attestation and is thus a good fit for enclaves.
\changed{
Moreover, as mentioned in the previous section, \pa must revoke group member keys in the event of a compromise.
EPID offers privacy-preserving \emph{signature-based revocation}, wherein the issuer can revoke any key using only a signature generated by that key.
Signature verifiers use signature revocation lists published by issuers to check whether the group member keys are revoked.
Using this mechanism, \sys provides \pa{}s with revocation capabilities without allowing them to link keys to individual users.
\pa{}s can define their own revocation policies to maximize their reputation and trustworthiness.
}

\subsection{Optimizations} \label{subsec:optm}
\subsubsection{Database Optimizations} \label{subsubsec:dboptm}
As with most modern database management systems, SQLite supports creating indexes in database tables to reduce query times.
Also, as discussed in Section \ref{sec:eval}, placing all timestamps for all servers in one table and conducting \texttt{JOIN} operations \changed{incurs} certain performance overhead.
An alternative is to use \changed{a separate table} per list.
However, we presented \sys evaluation results without creating any indexes or separate timestamp tables in order to show the worst-case performance.
Performance optimizations, such as changing the database layout, can be easily made by third parties, since they do not affect security of \sys.

\subsubsection{System-level Optimizations} \label{subsubsec:sysoptm}
As a system-level optimization, \sys can perform some processing steps in the background while waiting for the user to confirm the action, as follows: Server requests are sent to the enclave immediately, while the browser extension is displaying an approval notification.
As soon as the enclave receives the request, it runs the \texttt{get\_rate} function, while waiting for user approval before sending the rate-proof signature to the host and updating the MHT.
This optimization reduces user-perceived latency to that of client-side post-enclave and server-side EPID verification processes, essentially lowering total latency to $0.1751$ seconds -- 1/3 of the latency reported in Section~\ref{subsubsec:overall}.

\changed{
\subsubsection{Optimizing Pruning} \label{subsubsec:pruneoptm}
Although it is possible to create another \ecall for pruning, this might incur additional enclave entry/exit overhead (see Section \ref{subsec:dchal}).
Instead, pruning can be implemented within the \texttt{get\_rate} \ecall.
Since \texttt{get\_rate} always updates the hash chain and thus updates the MHT, we can prune if necessary to save processing time, rather than creating a new \ecall.
}

\subsection{Deploying \sys} \label{subsec:deploying}

\changed{
\subsubsection[Integration with CDNs and 3rd Party Providers]{Integration with CDNs and $3^{rd}$ Party Providers} \label{subsubsec:cdnintegration}
Although \sys aims to reduce developer efforts by choosing well-known primitives (e.g., SQLite and EPID), we do not expect all server operators to be experienced in implementing \sys components.
One common Web paradigm is Content Delivery Networks (CDNs) that are used reduce latency and operation costs.
CDNs have already recognized the opportunity to provide abuse prevention services to their customers.
For example, Cloudflare offers CAPTCHAs as a rate-limiting service~\cite{cfRateLimiting} to its wide user-base for free~\cite{cfhCaptcha}.
\sys could easily be adapted for use with CDNs, which would bring usability benefits across all websites served by the CDN.

Also, we expect independent (future) \sys providers to offer rate-proof services that are easy to integrate into websites -- similar to how CAPTCHAs are offered by reCAPTCHA~\cite{recaptcha} or hCAPTCHA~\cite{hCaptcha}.
These services would have to implement endpoints as discussed in Section~\ref{subsec:integration} and servers need to make minimal alterations to their websites.

\subsubsection{Website Operator Incentives} \label{subsubsec:websiteoperator}
Integrating \sys with a website is simple, as we have described in Section~\ref{subsec:deployability}.
There are several incentives for website operators to integrate \sys into their website.
First, usability of \sys.
\sys reduces CAPTCHA solving time from an average of $\approx 10$ seconds to a mere 0.25 seconds, thus drastically improves the user experience.
Second, privacy of \sys.
reCAPTCHAs are good in terms of usability, but have a concern regarding user privacy, as it uses long term state such as cookies.
Since \sys prevents generated rate-proofs to be linkable to a certain user, therefore is more private compared to reCAPTCHA systems.
Third, bandwidth of \sys.
The bandwidth consumption of \sys is an order of magnitude smaller compared to other CAPTCHA systems.
This helps save precious network bandwidth not only on the website operator side, but also for the user side.
These three points benefits users and may incentivize them to visit the website more often, connecting to website revenue via advertisement.

In fact, user demand for CAPTCHA solving time reduction and privacy has become significant that Cloudflare offers Privacy Pass~\cite{davidson2018privacy} aimed at reducing the number of CAPTCHAs showed to legitimate users -- especially useful for anonymous network and VPN users \cite{privacyPassCF}.

\subsubsection{\pa Operator Incentives} \label{subsubsec:paoperator}
Our \pa{s} are only responsible for initially issuing credentials to \sys enclaves.
Therefore, their workload is very lightweight.
They can be run by many organizations with different incentives, for example:
\begin{compactenum}
\item TEE hardware vendors wanting to increase the desirability of their hardware;
\item Online identity providers (e.g., Google, Facebook, Microsoft) who already provide services such as federated login;
\item For-profit businesses that charge a fee for premium support;
\item Non-profit organizations, similarly to the Let's Encrypt CA service.
\end{compactenum}
\sys users can, and are encouraged to, register with multiple \pa{s} and randomly select which private key to use for generating each rate-proof.
This allows new \pa{s} to join the \sys ecosystem and ensures that clients have maximum choice of \pa without the risk of vendor lock-in.
}

\subsubsection{Client-side components} \label{subsubsec:clientcomp}
On the client-side, \sys could be integrated into web browsers, and would thus work ``out of the box'' on platforms with a suitable TEE.

\section{Related Work} \label{sec:relw}
\changed{
\sys is situated in the intersection of multiple fields of research, including DoS (or Distributed DoS (DDoS)) protection, human presence, and CAPTCHA improvements and alternatives.
In this section, we discuss related work in each of these fields and their relevance to \sys.
}

\textbf{Network layer defenses.}
The main purpose of network layer DoS/DDoS protection mechanisms is to detect malicious network flows targeting the availability of the system.
This is done by using filtering~\cite{liu2008filter} or rate-limiting~\cite{cheng2002use} (or a combination thereof) according to certain characteristics of a flow.
We refer the reader to~\cite{peng2007survey} for an in-depth survey of network-level defenses.
Moreover, additional countermeasures can be employed depending on the properties of the system under attack (e.g., sensor-based networks~\cite{ouyang2011novel}, peer-to-peer networks~\cite{perlegos2004defense} and virtual ad-hoc networks~\cite{kerrache2017tfdd}).

\textbf{Application layer defenses.}
Application layer measures for DoS/DDoS protection focus on separating human-originated traffic from bot-originated traffic.
To this end, problems that are hard to solve by computers and (somewhat) easy to solve by humans comprise the basis of application layer solutions.
As explained in Section~\ref{sec:intro}, CAPTCHAs~\cite{VonAhn2003} are used extensively.
Although developing more efficient CAPTCHAs is an active area of research~\cite{gossweiler2009s, wang2011image, sanghavi2009progressive, datta2005imagination}, research aiming to subvert CAPTCHAs is also prevalent~\cite{mori2003recognizing, yan2008low, golle2008machine, gao2012divide}.
In addition to such automated attacks, CAPTCHAs suffer from inconsistency when solved by humans (e.g., perfect agreement when solved by three humans are 71\% and 31\% for image and audio CAPTCHAs, respectively~\cite{bursztein2010good}).
\cite{motoyama2010re} suggest that although CAPTCHAs succeed at telling humans and computers apart, by using CAPTCHA-solving services (operated by humans), with an acceptable cost, CAPTCHAs can be defeated.
Moreover, apart from questions regarding their efficacy, one other concern about CAPTCHAs is their usability.
Studies such as~\cite{fidas2011necessity, bursztein2010good} show that CAPTCHAs are not only difficult but also time-consuming for humans, with completion time of $\approx$10 seconds on average.
While behavioral CAPTCHAs are available, they suffer from privacy issues.
A prevalent example, reCAPTCHA~\cite{recaptcha}, works by analyzing user behavioral data (which requires sharing this data with the CAPTCHA provider) and claims to work more efficiently if used on multiple pages.
In contrast, \sys can provide at least the equivalent of abuse-prevention as CAPTCHAs, whilst minimizing the burden on users and offering strong privacy guarantees.

\textbf{Human presence detection.}
Human presence refers to determining whether specific actions were performed by a human.
VButton~\cite{li2018vbutton} proposes a system design based on ARM's TrustZone~\cite{holding2009arm}.
Secure detection of human presence is achieved by setting the display and the touch input peripherals as secure peripherals which can only be controlled by the TEE while VButton UI is displayed.
With a secure I/O mechanism in place, user actions can be authenticated to originate from VButton UI by a remote server using software attestation.
Similarly, Not-a-Bot~\cite{gummadi2009not} designs a system based on TPMs by tagging each network request with an attestation assuring that the request has been performed not long after a keyboard or mouse input by the user.
Unfortunately, Intel SGX does not support secure I/O and it is not currently possible to implement similar systems on devices with only Intel SGX support.
SGXIO~\cite{weiser2017sgxio} proposes an architecture for creating secure paths to I/O devices from enclaves using a trusted stack which contains a hypervisor, I/O drivers and an enclave for trusted boot.
In addition, an untrusted VM hosts secure applications.
The communication between secure applications and drivers are encrypted using keys generated at the end of the local attestation process.
Unfortunately, the implementation of this system is not yet available.
\changed{
Fidelius~\cite{fidelius} protects user secrets from a compromised browser or OS by protecting the path from the input and output peripherals to the hardware enclave.
Similarly to SGXIO, this is a promising step towards general-purpose trusted UI.
If trusted UI capabilities do become widely available on TEEs, these can complement our \sys design (e.g., providing stronger assurance of human presence).
}

\changed{\textbf{Privacy Pass.}
Privacy Pass~\cite{davidson2018privacy} implements a browser extension to reduce the burden of CAPTCHAs for legitimate users when visiting websites served by Cloudflare.
When a user solves a CAPTCHA, Cloudflare sends the user multiple anonymous cryptographic tokens, which the user can later ``spend'' to access Cloudflare-operated services without encountering additional CAPTCHAs Although Privacy Pass significantly benefits benign users, it could still be exploited by CAPTCHA farms.
Additionally, Privacy Pass' is currently limited to Cloudflare users.
}

\section{Conclusion \& Future Work} \label{sec:confw}
\sys is a novel approach for leveraging client-side TEEs to help legitimate clients avoid solving CAPTCHAs on the Web.
The unforgeable yet privacy-preserving rate-proofs generated by the TEE provide strong assurance that the client is not behaving abusively.
Our proof-of-concept implementation demonstrates that rate-proofs can be generated in less than $0.25$ seconds on commodity hardware, and that \sys reduces data transfer by more than $98\%$ compared to existing CAPTCHA schemes.
As for future work, we plan to employ optimization techniques discussed in Section~\ref{sec:disc}, \changed{implement and evaluate \sys{} on ARM TrustZone using OpenEnclave}, and explore new types of web security applications that are enabled using client-side TEEs.

\footnotesize
\raggedright
\bibliographystyle{abbrv}
\bibliography{references}

\begin{thebibliography}{10}

\bibitem{antiCaptcha}
Anticaptcha.
\newblock \url{https://anti-captcha.com/mainpage}, [Online] Accessed:
  2020-05-22.

\bibitem{nmprotocol}
Chrome native messaging protocol.
\newblock
  \url{https://developer.chrome.com/extensions/nativeMessaging#native-messaging-host-protocol},
  [Online] Accessed: 2020-02-09.

\bibitem{chromenotify}
Chrome notifications.
\newblock \url{https://developer.chrome.com/apps/notifications}, [Online]
  Accessed: 2020-02-14.

\bibitem{cfRateLimiting}
Cloudflare rate limiting.
\newblock \url{https://www.cloudflare.com/rate-limiting/}, [Online] Accessed:
  2020-05-19.

\bibitem{epidgithub}
Epid sdk.
\newblock \url{https://github.com/Intel-EPID-SDK/epid-sdk}, [Online] Accessed:
  2020-02-14.

\bibitem{chrome}
Google chrome.
\newblock \url{https://www.google.com/chrome/}, [Online] Accessed: 2020-02-11.

\bibitem{hCaptcha}
hcaptcha.
\newblock \url{https://www.hcaptcha.com/}, [Online] Accessed: 2020-05-21.

\bibitem{mc}
Intel dynamic application loader developer guide: Monotonic counters.
\newblock
  \url{https://software.intel.com/en-us/dal-developer-guide-features-monotonic-counters},
  [Online] Accessed: 2020-02-05.

\bibitem{ipp-optimized}
Intel integrated performance primitives cryptography.
\newblock \url{https://github.com/intel/ipp-crypto}, [Online] Accessed:
  2020-05-28.

\bibitem{nuc}
Intel nuc kit nuc7pjyh.
\newblock
  \url{https://ark.intel.com/content/www/us/en/ark/products/126137/intel-nuc-kit-nuc7pjyh.html},
  [Online] Accessed: 2020-02-11.

\bibitem{intelProcessor}
Intel® pentium® processor g4400.
\newblock
  \url{https://ark.intel.com/content/www/us/en/ark/products/88179/intel-pentium-processor-g4400-3m-cache-3-30-ghz.html},
  [Online] Accessed: 2020-05-19.

\bibitem{jsmn}
Jsmn json parser.
\newblock \url{https://github.com/zserge/jsmn}, [Online] Accessed: 2020-02-13.

\bibitem{mbedtlsgithub}
Mbed tls.
\newblock \url{https://github.com/ARMmbed/mbedtls}, [Online] Accessed:
  2020-02-14.

\bibitem{cfhCaptcha}
Moving from recaptcha to hcaptcha.
\newblock \url{https://blog.cloudflare.com/moving-from-recaptcha-to-hcaptcha/},
  [Online] Accessed: 2020-05-19.

\bibitem{nativeMessaging}
Native messaging.
\newblock \url{https://developer.chrome.com/extensions/nativeMessaging},
  [Online] Accessed: 2020-02-13.

\bibitem{oe}
Open enclave sdk.
\newblock \url{https://openenclave.io/sdk/}, [Online] Accessed: 2020-02-14.

\bibitem{dchestCaptcha}
Package captcha.
\newblock \url{https://github.com/dchest/captcha}, [Online] Accessed:
  2020-05-21.

\bibitem{recaptcha}
recaptcha.
\newblock \url{https://www.google.com/recaptcha/intro/v3.html}, [Online]
  Accessed: 2020-02-05.

\bibitem{recaptchadev}
recaptcha v2.
\newblock \url{https://developers.google.com/recaptcha/docs/display}, [Online]
  Accessed: 2020-02-13.

\bibitem{port}
runtime.port.
\newblock \url{https://developer.chrome.com/extensions/runtime#type-Port},
  [Online] Accessed: 2020-02-12.

\bibitem{svgCaptcha}
svg captcha.
\newblock \url{https://github.com/produck/svg-captcha}, [Online] Accessed:
  2020-05-21.

\bibitem{captchaSolverComp}
Top 10 captcha solving services compared.
\newblock
  \url{https://prowebscraper.com/blog/top-10-captcha-solving-services-compared/},
  [Online] Accessed: 2020-05-22.

\bibitem{privacyPassCF}
Using privacy pass with cloudflare.
\newblock
  \url{https://support.cloudflare.com/hc/en-us/articles/115001992652-Using-Privacy-Pass-with-Cloudflare},
  [Online] Accessed: 2020-06-01.

\bibitem{anati2013innovative}
I.~Anati, S.~Gueron, S.~Johnson, and V.~Scarlata.
\newblock Innovative technology for cpu based attestation and sealing.
\newblock In {\em Proceedings of the 2nd international workshop on hardware and
  architectural support for security and privacy}, volume~13, page~7. ACM New
  York, NY, USA, 2013.

\bibitem{Ateniese2000group}
G.~Ateniese, J.~Camenisch, M.~Joye, and G.~Tsudik.
\newblock {A Practical and Provably Secure Coalition-Resistant Group Signature
  Scheme}.
\newblock In M.~Bellare, editor, {\em Advances in Cryptology --- CRYPTO 2000},
  pages 255--270, Berlin, Heidelberg, 2000. Springer Berlin Heidelberg.

\bibitem{au2006constant}
M.~H. Au, W.~Susilo, and Y.~Mu.
\newblock Constant-size dynamic k-taa.
\newblock In {\em International conference on security and cryptography for
  networks}, pages 111--125. Springer, 2006.

\bibitem{brickell2004direct}
E.~Brickell, J.~Camenisch, and L.~Chen.
\newblock Direct anonymous attestation.
\newblock In {\em Proceedings of the 11th ACM conference on Computer and
  communications security}, pages 132--145, 2004.

\bibitem{brickell2012static}
E.~Brickell, L.~Chen, and J.~Li.
\newblock A static diffie-hellman attack on several direct anonymous
  attestation schemes.
\newblock In {\em International Conference on Trusted Systems}, pages 95--111.
  Springer, 2012.

\bibitem{epid}
E.~Brickell and J.~Li.
\newblock Enhanced privacy id: A direct anonymous attestation scheme with
  enhanced revocation capabilities.
\newblock In {\em Proceedings of the 2007 ACM Workshop on Privacy in Electronic
  Society}, WPES ’07, page 21–30, New York, NY, USA, 2007. Association for
  Computing Machinery.
\newblock \url{https://doi.org/10.1145/1314333.1314337}.

\bibitem{bursztein2010good}
E.~Bursztein, S.~Bethard, C.~Fabry, J.~C. Mitchell, and D.~Jurafsky.
\newblock How good are humans at solving captchas? a large scale evaluation.
\newblock In {\em 2010 IEEE symposium on security and privacy}, pages 399--413.
  IEEE, 2010.

\bibitem{cheng2002use}
C.-M. Cheng, H.~Kung, and K.-S. Tan.
\newblock Use of spectral analysis in defense against dos attacks.
\newblock In {\em Global Telecommunications Conference, 2002. GLOBECOM'02.
  IEEE}, volume~3, pages 2143--2148. IEEE, 2002.

\bibitem{AndroidProtectedConfirmation}
J.~Danisevskis.
\newblock Android protected confirmation: Taking transaction security to the
  next level.
\newblock
  \url{https://developer.android.com/training/articles/security-android-protected-confirmation},
  [Online] Accessed: 2020-02-05.

\bibitem{datta2005imagination}
R.~Datta, J.~Li, and J.~Z. Wang.
\newblock Imagination: a robust image-based captcha generation system.
\newblock In {\em Proceedings of the 13th annual ACM international conference
  on Multimedia}, pages 331--334, 2005.

\bibitem{davidson2018privacy}
A.~Davidson, I.~Goldberg, N.~Sullivan, G.~Tankersley, and F.~Valsorda.
\newblock Privacy pass: Bypassing internet challenges anonymously.
\newblock {\em Proceedings on Privacy Enhancing Technologies},
  2018(3):164--180, 2018.

\bibitem{initpresenceattestation}
X.~Ding and G.~Tsudik.
\newblock Initializing trust in smart devices via presence attestation.
\newblock {\em Computer Communications}, 131:35 -- 38, 2018.
\newblock
  \url{http://www.sciencedirect.com/science/article/pii/S0140366418304882}.

\bibitem{fidelius}
S.~{Eskandarian}, J.~{Cogan}, S.~{Birnbaum}, P.~C.~W. {Brandon}, D.~{Franke},
  F.~{Fraser}, G.~{Garcia}, E.~{Gong}, H.~T. {Nguyen}, T.~K. {Sethi},
  V.~{Subbiah}, M.~{Backes}, G.~{Pellegrino}, and D.~{Boneh}.
\newblock Fidelius: Protecting user secrets from compromised browsers.
\newblock In {\em 2019 IEEE Symposium on Security and Privacy (SP)}, pages
  264--280, 2019.

\bibitem{fidas2011necessity}
C.~A. Fidas, A.~G. Voyiatzis, and N.~M. Avouris.
\newblock On the necessity of user-friendly captcha.
\newblock In {\em Proceedings of the SIGCHI Conference on Human Factors in
  Computing Systems}, pages 2623--2626, 2011.

\bibitem{gao2012divide}
H.~Gao, W.~Wang, and Y.~Fan.
\newblock Divide and conquer: an efficient attack on yahoo! captcha.
\newblock In {\em 2012 IEEE 11th International Conference on Trust, Security
  and Privacy in Computing and Communications}, pages 9--16. IEEE, 2012.

\bibitem{golle2008machine}
P.~Golle.
\newblock Machine learning attacks against the asirra captcha.
\newblock In {\em Proceedings of the 15th ACM conference on Computer and
  communications security}, pages 535--542, 2008.

\bibitem{gossweiler2009s}
R.~Gossweiler, M.~Kamvar, and S.~Baluja.
\newblock What's up captcha? a captcha based on image orientation.
\newblock In {\em Proceedings of the 18th international conference on World
  wide web}, pages 841--850, 2009.

\bibitem{gummadi2009not}
R.~Gummadi, H.~Balakrishnan, P.~Maniatis, and S.~Ratnasamy.
\newblock Not-a-bot: Improving service availability in the face of botnet
  attacks.
\newblock In {\em NSDI}, volume~9, pages 307--320, 2009.

\bibitem{hoekstra2013using}
M.~Hoekstra, R.~Lal, P.~Pappachan, V.~Phegade, and J.~Del~Cuvillo.
\newblock Using innovative instructions to create trustworthy software
  solutions.
\newblock {\em HASP@ ISCA}, 11(10.1145):2487726--2488370, 2013.

\bibitem{holding2009arm}
A.~Holding.
\newblock Arm security technology, building a secure system using trustzone
  technology, 2009.

\bibitem{kerrache2017tfdd}
C.~A. Kerrache, N.~Lagraa, C.~T. Calafate, and A.~Lakas.
\newblock Tfdd: A trust-based framework for reliable data delivery and dos
  defense in vanets.
\newblock {\em Vehicular Communications}, 9:254--267, 2017.

\bibitem{leung2008possible}
A.~Leung, L.~Chen, and C.~J. Mitchell.
\newblock On a possible privacy flaw in direct anonymous attestation (daa).
\newblock In {\em International Conference on Trusted Computing}, pages
  179--190. Springer, 2008.

\bibitem{li2018vbutton}
W.~Li, S.~Luo, Z.~Sun, Y.~Xia, L.~Lu, H.~Chen, B.~Zang, and H.~Guan.
\newblock Vbutton: Practical attestation of user-driven operations in mobile
  apps.
\newblock In {\em Proceedings of the 16th Annual International Conference on
  Mobile Systems, Applications, and Services}, pages 28--40, 2018.

\bibitem{liu2008filter}
X.~Liu, X.~Yang, and Y.~Lu.
\newblock To filter or to authorize: Network-layer dos defense against
  multimillion-node botnets.
\newblock In {\em Proceedings of the ACM SIGCOMM 2008 conference on Data
  communication}, pages 195--206, 2008.

\bibitem{mckeen2013innovative}
F.~McKeen, I.~Alexandrovich, A.~Berenzon, C.~V. Rozas, H.~Shafi, V.~Shanbhogue,
  and U.~R. Savagaonkar.
\newblock Innovative instructions and software model for isolated execution.
\newblock {\em Hasp@ isca}, 10(1), 2013.

\bibitem{mori2003recognizing}
G.~Mori and J.~Malik.
\newblock Recognizing objects in adversarial clutter: Breaking a visual
  captcha.
\newblock In {\em 2003 IEEE Computer Society Conference on Computer Vision and
  Pattern Recognition, 2003. Proceedings.}, volume~1, pages I--I. IEEE, 2003.

\bibitem{motoyama2010re}
M.~Motoyama, K.~Levchenko, C.~Kanich, D.~McCoy, G.~M. Voelker, and S.~Savage.
\newblock Re: Captchas-understanding captcha-solving services in an economic
  context.
\newblock In {\em USENIX Security Symposium}, volume~10, page~3, 2010.

\bibitem{ouyang2011novel}
X.~Ouyang, B.~Tian, Q.~Li, J.-y. Zhang, Z.-M. Hu, and Y.~Xin.
\newblock A novel framework of defense system against dos attacks in wireless
  sensor networks.
\newblock In {\em 2011 7th International Conference on Wireless Communications,
  Networking and Mobile Computing}, pages 1--5. IEEE, 2011.

\bibitem{peng2007survey}
T.~Peng, C.~Leckie, and K.~Ramamohanarao.
\newblock Survey of network-based defense mechanisms countering the dos and
  ddos problems.
\newblock {\em ACM Computing Surveys (CSUR)}, 39(1):3--es, 2007.

\bibitem{perlegos2004defense}
P.~Perlegos.
\newblock {\em DoS defense in structured peer-to-peer networks}.
\newblock Computer Science Division, University of California, 2004.

\bibitem{enclavedb}
C.~Priebe, K.~Vaswani, and M.~Costa.
\newblock Enclavedb: A secure database using sgx.
\newblock In {\em 2018 IEEE Symposium on Security and Privacy (SP)}, pages
  264--278. IEEE, 2018.

\bibitem{cryptoeprint:2019:757}
J.~Protzenko, B.~Parno, A.~Fromherz, C.~Hawblitzel, M.~Polubelova,
  K.~Bhargavan, B.~Beurdouche, J.~Choi, A.~Delignat-Lavaud, C.~Fournet,
  N.~Kulatova, T.~Ramananandro, A.~Rastogi, N.~Swamy, C.~Wintersteiger, and
  S.~Zanella-Beguelin.
\newblock Evercrypt: A fast, verified, cross-platform cryptographic provider.
\newblock Cryptology ePrint Archive, Report 2019/757, 2019.
\newblock \url{https://eprint.iacr.org/2019/757}.

\bibitem{rudolph2007covert}
C.~Rudolph.
\newblock Covert identity information in direct anonymous attestation (daa).
\newblock In {\em IFIP International Information Security Conference}, pages
  443--448. Springer, 2007.

\bibitem{sanghavi2009progressive}
M.~Sanghavi and S.~Doshi.
\newblock Progressive captcha, Apr.~30 2009.
\newblock US Patent App. 11/929,716.

\bibitem{VonAhn2003}
L.~von Ahn, M.~Blum, N.~J. Hopper, and J.~Langford.
\newblock {CAPTCHA: Using Hard AI Problems for Security}.
\newblock In E.~Biham, editor, {\em Advances in Cryptology --- EUROCRYPT 2003},
  pages 294--311, Berlin, Heidelberg, 2003. Springer Berlin Heidelberg.

\bibitem{wang2011image}
J.~Z. Wang, R.~Datta, and J.~Li.
\newblock Image-based captcha generation system, Apr.~19 2011.
\newblock US Patent 7,929,805.

\bibitem{weiser2017sgxio}
S.~Weiser and M.~Werner.
\newblock Sgxio: Generic trusted i/o path for intel sgx.
\newblock In {\em Proceedings of the Seventh ACM on Conference on Data and
  Application Security and Privacy}, pages 261--268, 2017.

\bibitem{yan2008low}
J.~Yan and A.~S. El~Ahmad.
\newblock A low-cost attack on a microsoft captcha.
\newblock In {\em Proceedings of the 15th ACM conference on Computer and
  communications security}, pages 543--554, 2008.

\bibitem{presenceattestation}
Z.~Zhang, X.~Ding, G.~Tsudik, J.~Cui, and Z.~Li.
\newblock Presence attestation: The missing link in dynamic trust
  bootstrapping.
\newblock In {\em Proceedings of the 2017 ACM SIGSAC Conference on Computer and
  Communications Security}, CCS ’17, page 89–102, New York, NY, USA, 2017.
  Association for Computing Machinery.
\newblock \url{https://doi.org/10.1145/3133956.3134094}.

\end{thebibliography}

\balance

\normalsize

\end{document}